\documentclass[acmsmall,screen]{acmart}
\AtBeginDocument{
  }
\usepackage{diagbox}
\usepackage{multirow}
\usepackage{setspace}
\usepackage{algorithm}
\usepackage{algorithmic}
\usepackage{pifont}
\usepackage{booktabs}
\usepackage{enumitem}
\usepackage{wrapfig}
\usepackage{graphicx}
\usepackage{textcomp}
\usepackage{colortbl}
\definecolor{c1}{HTML}{1F77B4}
\definecolor{c2}{HTML}{F5F5F5}
\usepackage{color}
\usepackage{tabularx}
\usepackage{tabulary}
\usepackage{caption}
\usepackage{subcaption}
\usepackage{rotating}
\usepackage{wrapfig}
\usepackage{threeparttable}
\usepackage[misc]{ifsym}
\usepackage{array}
\usepackage{microtype}
\setcopyright{acmlicensed}
\copyrightyear{2025}
\acmYear{2025}
\acmDOI{XXXXXXX.XXXXXXX}

\begin{document}

\title{Rethinking LSM-tree based Key-Value Stores: A Survey}

\author{Yina Lv}
\email{elainelv95@gmail.com}
\orcid{0000-0003-3971-3123}
\affiliation{
  \institution{School of Informatics, Xiamen University}
  \country{China}
}

\author{Qiao Li}
\authornote{Corresponding Authors}
\email{qiaoli045@gmail.com}
\orcid{0000-0002-4579-4268}
\affiliation{
  \institution{Department of Computer Science, Mohamed bin Zayed University of Artificial Intelligence, Abu Dhabi}\country{UAE}
}

\author{Quanqing Xu}
\authornotemark[1]
\email{xuquanqing.xqq@oceanbase.com}
\orcid{0000-0001-8989-9662}
\affiliation{
  \institution{OceanBase, Ant Group}\country{China}
}
\author{Congming Gao}
\email{gaocm@xmu.edu.cn}
\orcid{0000-0003-2611-2652}
\affiliation{
  \institution{School of Informatics, Xiamen University}\country{China}
}
\author{Chuanhui Yang}
\email{rizhao.ych@oceanbase.com}
\orcid{0009-0009-3530-6476}
\affiliation{
  \institution{OceanBase, Ant Group}\country{China}
}
\author{Xiaoli Wang}
\email{xlwang@xmu.edu.cn}
\orcid{0000-0002-8677-9080}
\affiliation{
  \institution{School of Informatics, Xiamen University}\country{China}
}
\author{Chun Jason Xue}
\email{Jason.xue@mbzuai.ac.ae}
\orcid{0000-0002-6431-9868}
\affiliation{
  \institution{Department of Computer Science, Mohamed bin Zayed University of Artificial Intelligence, Abu Dhabi}\country{UAE}
}
\renewcommand{\shortauthors}{Yina Lv, Qiao Li, et al.} 

\begin{abstract}
LSM-tree is a widely adopted data structure in modern key-value store systems that optimizes write performance in write-heavy applications by using append writes to achieve sequential writes. However, the unpredictability of LSM-tree compaction introduces significant challenges, including performance variability during peak workloads and in resource-constrained environments, write amplification caused by data rewriting during compactions, read amplification from multi-level queries, trade-off between read and write performance, as well as efficient space utilization to mitigate space amplification. Prior studies on LSM-tree optimizations have addressed the above challenges; however, in recent years, research on LSM-tree optimization has continued to propose. The goal of this survey is to review LSM-tree optimization, focusing on representative works in the past five years. This survey first studies existing solutions on how to mitigate the performance impact of LSM-tree flush and compaction and how to improve basic key-value operations. In addition, distributed key-value stores serve multi-tenants, ranging from tens of thousands to millions of users with diverse requirements. We then analyze the new challenges and opportunities in these modern architectures and across various application scenarios. Unlike the existing survey papers, this survey provides a detailed discussion of the state-of-the-art work on LSM-tree optimizations and gives future research directions.
\end{abstract}





\maketitle

% \tableofcontents

% Google BigTable (2006)
% Amazon Dynamo (2007)
% Apache Cassandra (2008)
% Apache HBase (2010)
% Google LevelDB (2011)
% Facebook RocksDB (2013)
% PingCAP TiKV (2018)
\section{Introduction}\label{sec:intro}
The log-structured merge-tree (LSM-tree) \cite{O1996LSMTree} has emerged as a mainstream storage engine in modern local key-value stores (i.e., Google LevelDB \cite{LevelDB}, Facebook RocksDB \cite{RocksDB}, etc) and distributed key-value stores (i.e., Google BigTable \cite{TOCS2008Bigtable}, Amazon Dynamo \cite{SOSP2007Dynamo}, Apache Cassandra \cite{Cassandra}, Apache HBase \cite{HBase}, PingCAP TiKV \cite{TiKV}, Alibaba OceanBase \cite{VLDB2022oceanbase,VLDB2023oceanbase}, etc), primarily due to its good capability to enhance write performance, especially in write-heavy scenarios like modern large language models (LLMs) with large-scale dataset scenarios \cite{ASPLOS2023Cooperative}.
The compaction process intrinsic to LSM-trees is essential for maintaining read performance and releasing storage space in \underline{LSM-tree based key-value stores} (\textbf{LSM-KVS}).
However, compaction operations are resource-intensive, consuming substantial CPU, I/O bandwidth, and memory resources.
This often suffers from severe resource contention between foreground user operations and background flush/compactions, resulting in a \textit{write-stall issue} \cite{VLDB2019performance} that the writing process may be slowed down or even stopped, especially when the system experiences multiple concurrent internal operations \cite{ATC2019SILK,VLDB2021constructing,FAST2023ADOC,HotStorage2024ELMo-Tune,SIGMOD2025RethinkingCompaction}.
Furthermore, this unpredictability of the compaction process poses significant challenges, particularly in performance variability during peak workloads and in systems with constrained resources, rewrites data-induced write amplification issues due to aggressive compactions, multiple-level query-induced read amplification issues due to lazy compactions, the trade-off between read and write performance, as well as efficient space management to prevent space amplification issues.
The challenges are further exacerbated by evolving hardware architectures and diverse application requirements that require specialized LSM-tree implementations.

Existing works have focused on LSM-tree optimizations from different perspectives.
This paper provides a comprehensive review and discussion of the existing works for LSM-tree optimizations, including designs of compactions and key-value store operations, optimizations for emerging architectures, and complex applications.
Specifically, we first review various internal compaction designs, including resource allocation between key-value store operations and internal flush/compaction operations, prioritizing the different operations on demand, and refining the compaction strategies to achieve superior performance.
Second, we review the optimization of key-value store operations, including point lookup, range query operations, and privacy-related deletes.
Furthermore, we analyze the design challenges of LSM-tree in distributed key-value stores, in-storage compaction designs, emerging hardware designs, and heterogeneous storage devices.
We also discuss real user cases from the application perspective, including workload-aware LSM-tree designs, tailored LSM-tree designs, and LSM-tree designs in multi-tenant scenarios.

\paragraph{\textbf{Contributions of This Survey.}}
The LSM-tree structure was proposed over 20 years ago and has been widely adopted in various key-value stores, ranging from single-node to distributed architectures.
However, real-world scenarios offer significant potential for further optimization, particularly when focusing on compute-storage disaggregation, multi-tenant environments, emerging hardware designs, and the use of heterogeneous storage devices.
The evolution of these trends has introduced both new challenges and opportunities in the management of LSM-tree architectures.
This survey aims to provide an overview of existing solutions, enabling readers to gain a deeper understanding of the issues in LSM-tree designs and to identify potential research directions.

The main contributions are listed as follows.
\begin{itemize}
\item We provide a comprehensive review of the research questions and designs of LSM-KVS, and discuss existing solutions (especially those from the past five years, covering more than 100 papers from 2020 to 2025) that aim to mitigate the impact of LSM-tree compactions on system performance and enhance the management of LSM-tree basic operations.
\item We make an in-depth discussion of both the challenges and opportunities presented by emerging hardware architectures and their diverse applications.
\item We outline key research directions and future trends in LSM-KVS, offering insights for optimizing LSM-tree designs.
\end{itemize}

\paragraph{\textbf{Comparisons of Existing Surveys in LSM-trees.}}
Through extensive studies, we have found that while there are some surveys related to the LSM-tree designs, these existing survey papers either have been published earlier or are proposed for specific optimization goals, thus lacking a comprehensive review of LSM-trees in emerging hardware, complex applications, and new architectures.
Luo et al. \cite{VLDBJ2020lsmbased}, published in 2020, classify and discuss existing work in write amplification, merge operations, hardware, special workloads, auto-tuning, and secondary indexing.
Some other survey papers \cite{2024Critical,mishra2024survey} only cited a limited number of papers, lacking a comprehensive analysis of the research in the field. 
For example, Vasileios et al. \cite{2024Critical} provided a survey on LSM-trees, which included only 12 references.
Mishra \cite{mishra2024survey} introduced the basics of LSM-tree and its integration with non-volatile memory; however, only eight papers were compared and discussed.
In addition, some works \cite{VLDB2021constructing,SIGMOD2023SplinterDB,SIGMOD2025RethinkingCompaction,VLDB2022Endure,VLDBJ2024towards,SIGMOD2024Moose} are not survey papers but provide optimization from various perspectives.
For example, some works \cite{VLDB2021constructing,SIGMOD2023SplinterDB,SIGMOD2025RethinkingCompaction}
are related to compaction strategy optimization, discussing the compaction optimization from four aspects, including when to trigger compaction, selecting which data and how much data during compaction, and data layout.
Endure \cite{VLDB2022Endure,VLDBJ2024towards} focused on the impact of compaction strategy, size ratio, and memory allocation choices on query performance, and considers the characteristics of the workload to optimize for maximizing worst-case throughput.
Moose \cite{SIGMOD2024Moose} addressed the challenges of optimizing point lookup, range lookup, and update operations by allowing independent configuration adjustments per level, revealing critical insights for performance optimization.
Unlike existing works, this survey paper provides an in-depth review of LSM-KVS, specifically focusing on works published over the past five years.

The paper is organized as follows.
Section \ref{sec:background} introduces LSM-KVS basics.
Section \ref{sec:review} reviews research challenges and state-of-the-art solutions in LSM-KVS design.
Section \ref{sec:application} discusses application-driven optimizations.
Section \ref{sec:architecture} analyzes architectural trade-offs and design goals.
Section \ref{sec:future} outlines research directions and future trends in LSM-KVS.
Finally, Section \ref{sec:conclusion} summarizes key insights.

\section{Background and Motivation}\label{sec:background}
\subsection{LSM-tree based Key-Value Store (LSM-KVS)}\label{sec:LSM-tree}
\begin{figure*}[t]
\centering
\includegraphics[width=0.98\textwidth]{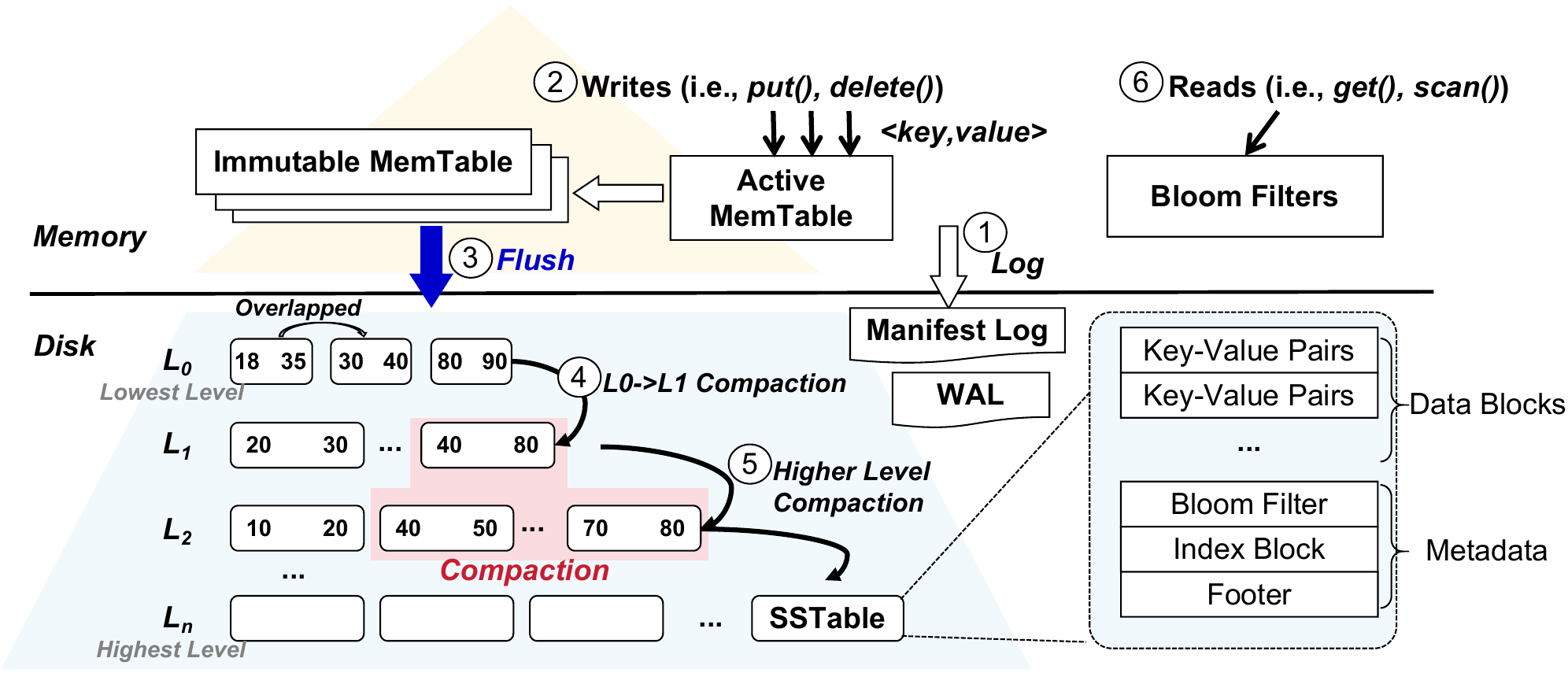}
\vspace{-0.1in}
\caption{The Architecture of LSM-KVS and Internal Management.}
\label{fig:LSMtree}
\vspace{-0.1in}
\end{figure*}

LSM-KVS is an efficient data storage architecture specifically designed for key-value storage scenarios \cite{SIGMOD2020KVSEngines,FAST2022Closing,TCAD2022LLSM,ICDE2024KD}. 
This architecture organizes data as key-value pairs, where each key serves as a unique identifier for its corresponding value, which contains the actual data.
These key-value pairs are managed by the LSM-tree, which has a log-structured feature to achieve high performance in write-intensive applications.
This design enables the system to efficiently handle massive write operations by first recording them sequentially in an append-only log.
Subsequently, during periods of lower system activity, these changes are merged with existing data through a background compaction process.
To do this, we can transform random write operations into sequential I/O operations, significantly improving performance as sequential I/O is typically an order of magnitude faster than random I/O in modern storage systems.
The adoption of the LSM-KVS architecture optimizes write speed and storage efficiency, making it an ideal choice for write-intensive applications, such as real-time analysis, stream processing, time-series databases, logging archive systems, and cloud storage services.
Figure \ref{fig:LSMtree} illustrates an overview of the LSM-KVS architecture and its internal management, including the MemTables in memory and SSTables on disk.
In the following, we will provide a detailed discussion of LSM-KVS architecture and its internal management.

\subsubsection{LSM-KVS Architecture}
The main components of the LSM-KVS architecture include MemTable, SSTable, write-ahead log, and Bloom filter for the SSTable.
In the following, we will introduce these components one by one in detail.

\textbf{MemTable (Memory Table).}
LSM-tree write operations (i.e., inserts, updates, deletes) are performed in an out-of-place manner, where the MemTable in the memory space serves as the staging area to absorb recent writes before they are persisted to disk.
In this case, MemTable acts as an in-memory buffer, providing a fast process for newly inserted or updated data and ensuring low-latency responses to user operations.
The MemTable is designed to handle high-throughput write operations efficiently, utilizing a compact data structure that optimizes both space efficiency and data retrieval speed.
When the MemTable reaches a predefined capacity threshold (default 64MB in RocksDB \cite{RocksDB}), it stops being written and becomes immutable.
Then, a new MemTable takes over as the active buffer for incoming writes.
The immutable MemTable is then written to disk in a process known as \textit{flushing}, which transfers the data to the persistent storage level within the LSM-tree, ensuring data durability.
The MemTable design is not only central to the ability of LSM-tree to handle write-intensive workloads, but also plays a significant role in read performance.
Since recently accessed data are likely to be in the MemTable, read operations can often be satisfied directly from memory, bypassing the slower disk-based lookups or queries.

\textbf{SSTable (Sorted String Table).}
Upon being written to disk, the MemTable is transformed into an SSTable, essentially a snapshot of the MemTable at the time of flushing.
The SSTable is the foundational on-disk data structure, which is immutable.
Any updates to the data result in new versions of the related SSTables in an out-of-place manner.
This immutability of SSTables simplifies the management of concurrent reads and writes, since read operations can access SSTables without locking, and write operations can proceed without interfering with ongoing reads.
As illustrated in Figure \ref{fig:LSMtree}, the internal structure of an SSTable typically consists of multiple data blocks that store the actual sorted \textit{key-value pairs}, an index block that provides pointers to the data blocks for quick data retrieval, and a Bloom filter to quickly check the presence of a key.
The data blocks may also be compressed to save storage space.
The entire structure of the SSTable is designed to support efficient data retrieval and maintenance of data persistence and consistency within the LSM-tree.
In addition, due to their sorted nature, they support fast range queries, which is particularly beneficial for applications that require scanning large datasets.

\textbf{WAL (Write-Ahead Log).}
The WAL protocol follows the ``write-ahead'' principle, requiring that any modification to the key-value store be recorded in the log before being committed to the actual data file (\ding{172}).
WAL ensures data durability and consistency in the event of a system crash or power failure, which is particularly important for providing a reliable recovery mechanism.
WAL also contributes to performance optimization. 
By batching multiple log entries together, the system can reduce the number of disk writes required, thereby improving write throughput.
Moreover, WAL allows for the concurrent execution of transactions without interfering with each other, as each transaction's changes are first recorded in the log, ensuring that even if one transaction fails, it does not impact the others.
In addition, there is also a manifest log that tracks metadata changes.

\textbf{Bloom Filter.}
Bloom filters are space-efficient probabilistic data structures that efficiently test set elements, offering strong guarantees against false negatives while maintaining a tunable \textit{false positive rate}.
The Bloom filter is composed of an $m$-bit array (initialized to zero) and $k$ independent hash functions, which involves two operations: insertion and query.
During insertion, an element $x$ is processed by computing its $k$ hash values $\{h_1(x), \ldots, h_k(x)\}$ and setting the corresponding array bits to one.
When querying for $x$, the same hash functions are reapplied to check the relevant bit positions.
If any bit remains unset, $x$ is definitively not in the set (guaranteeing no false negatives); if all bits are set, $x$ is probably present with false positive probability $\epsilon \approx (1-e^{-kn/m})^k$, where $n$ denotes the number of inserted elements.
This probability can be optimized by selecting $k \approx (m/n)\ln 2$ hash functions, creating an effective tradeoff between space overhead ($m$) and computational cost ($k$).
Note that a Bloom filter can never definitively confirm the presence of an element.
The false positive rate is influenced by the bit array size and the number of hash functions: More hash functions can reduce errors, but at the cost of space and computation.
Despite their probabilistic nature, Bloom filters are valuable in key-value stores because they quickly rule out non-existent items and minimize expensive disk or data retrieval operations.

\textbf{Key-Value Store Operations.}
Based on the above main components, LSM-trees support various key-value store operations, such as insert/update operations (i.e., $put()$), point lookup and range query operations (i.e., $get()$, $scan()$), as well as delete operations (i.e., $delete()$).
During the processing of write operations, we insert new data or update existing data by issuing a $put()$ command (\ding{173}).
This operation first writes the data to the MemTable, which stores recent data changes.
To retrieve data, users perform a $get()$ or $scan()$ command (\ding{177}).
The system first checks the MemTable for the most recent version of the data.
If the data is not found in the MemTable, the system then searches through the SSTables on disk, starting with the lowest level (most recent) and moving upwards until the key is matched.
This process often leverages in-memory data structures, such as the Bloom filter, to optimize the performance.
Unlike lookup operations, range queries retrieve a set of keys within a certain range.
This operation may require scanning multiple SSTables and merging the latest version for each required entry, as the data is sorted but not necessarily indexed by range.
When a query is executed, only the latest version of the data is returned, and all previous versions, which have been logically invalidated by subsequent updates, are discarded.
When a delete command is issued for a specific key, the key-value store records this as a deletion marker, often in the form of a $tombstone$, within the MemTable or the corresponding SSTable.
This logical invalidation ensures that subsequent queries do not return the outdated data, even though the physical data remains on disk until a compaction occurs.
Compaction is the background process responsible for merging SSTables and physically removing marked data. 
A deletion is ultimately regarded as persistent only after its associated $tombstone$ has been propagated to the final level of the LSM-tree. 
At this time, the correlated $tombstone$ can be removed.
These operations in the key-value stores highlight the unique characteristics of LSM-trees, which prioritize write performance and are optimized for scenarios where data is written more frequently than read.
This structure allows LSM-KVS to handle large volumes of data while maintaining high throughput and efficient storage utilization.

\subsubsection{Compactions in LSM-trees}
The efficiency of LSM-tree management is significantly influenced by \textit{how it handles internal flush and compaction operations}, which are crucial for merging data across different levels of the LSM-tree to ensure read performance and optimize storage utilization.

\textbf{MemTable to SSTable Flush.} 
The initial layer of the LSM-tree, MemTable, serves as a high-speed buffer for recent writes.
Given its in-memory nature, MemTable offers rapid access but is volatile.
To ensure data durability, the immutable MemTable will be flushed to disk (\ding{174}), becoming a new SSTable at the highest level of the LSM-tree, referred to as $L_0$.
This flush operation is a critical process of LSM-tree management, as it transforms volatile data into a persistent state.
The timing of flush operations is decided by the allocated memory capacity or the parameter set by the database manager.
Following the flush process, the MemTable is cleared from memory, releasing space for subsequent write operations and avoiding memory overflow.
Although flush operations are essential background I/O tasks that manage data durability, they can potentially impact ongoing key-value store operations, affecting overall system performance.
This is because the I/O bandwidth and resources are shared between the flush operations and the key-value store operations, which can lead to contention and reduced throughput.

\textbf{Compaction Triggers.} 
As more data is written, the number of SSTables at each level increases.
To prevent excessive read amplification and maintain efficient data retrieval, compactions are triggered among the different levels (i.e., $L_0$-$L_n$) of the LSM-tree in the disk.
Take the RocksDB as an example, as illustrated in Figure \ref{fig:LSMtree}.
Level 0, referred to as $L_0$, is organized in tiering, and other levels are organized in leveling.
Within $L_0$, SSTables may overlap with others because the data is directly flushed from memory.
Some LSM-tree-based systems execute intra-level compactions within $L_0$, such as OceanBase \cite{VLDB2022oceanbase,VLDB2023oceanbase}.
When the number of SSTables exceeds a certain threshold or the system receives the command from the database manager, the compactions from $L_0$ to $L_1$ will be triggered to collect the victim SSTables (\ding{175}).
Higher-level compactions involve merging SSTables across multiple levels (from $L_1$ to $L_n$), which are important for reclaiming space from updated or deleted data (\ding{176}).
Compared with low-level compactions, high-level compactions are more resource-intensive and are scheduled based on size or count thresholds, time intervals, or system-defined policies.
The above compaction operations will occur when the number of flush operations exceeds a certain threshold or are manually executed at idle time.
During the execution of compaction operations, an SSTable is selected as a victim SSTable based on the compaction policies.
Then, the overlapped SSTables in the next level will be read together into memory, and these SSTables will be merged with the victim SSTable.
It reads the key-value pairs from these SSTables and sorts them to ensure that they are in the correct order.
Duplicate keys are identified and deleted, usually by keeping the most recent version of the data.
Obsolete data, such as $tombstones$ or updated values, are removed during this process.
After compaction, the sorted and cleaned data is then written back to new SSTables, which replace the old ones.
As we can see, compaction ensures the sequentiality of data.
However, this process involves additional reads and writes because the management of the LSM-tree triggers them and constant writes from the foreground workloads, which may seriously conflict with key-value store operations.

In conclusion, the design and optimization of LSM-tree-based storage engines requires simultaneously considering high write throughput by minimizing I/O bottlenecks during internal flush and SSTable merges, high read performance through efficient indexing and Bloom filter utilization, and adapting dynamically to workload patterns and application requirements.

\subsection{Distributed LSM-KVS}\label{sec:LSM-KVS}
Distributed LSM-KVS architecture has emerged as a key solution, addressing the limitations of traditional single-node systems (e.g., LevelDB, RocksDB) in hyperscale data center environments. 
Whereas conventional single-node designs excel at local storage optimization through compaction and caching mechanisms, they lack distributed coordination primitives required for geo-distributed deployments.
Modern distributed LSM-KVS implementations overcome these limitations by introducing sharding-aware data placement strategies, global resource orchestration, and data replication, collectively enabling linear scalability, fault tolerance across availability zones, and workload-aware performance optimization.

\begin{figure*}[htbp]
\centering
\includegraphics[width=0.92\textwidth]{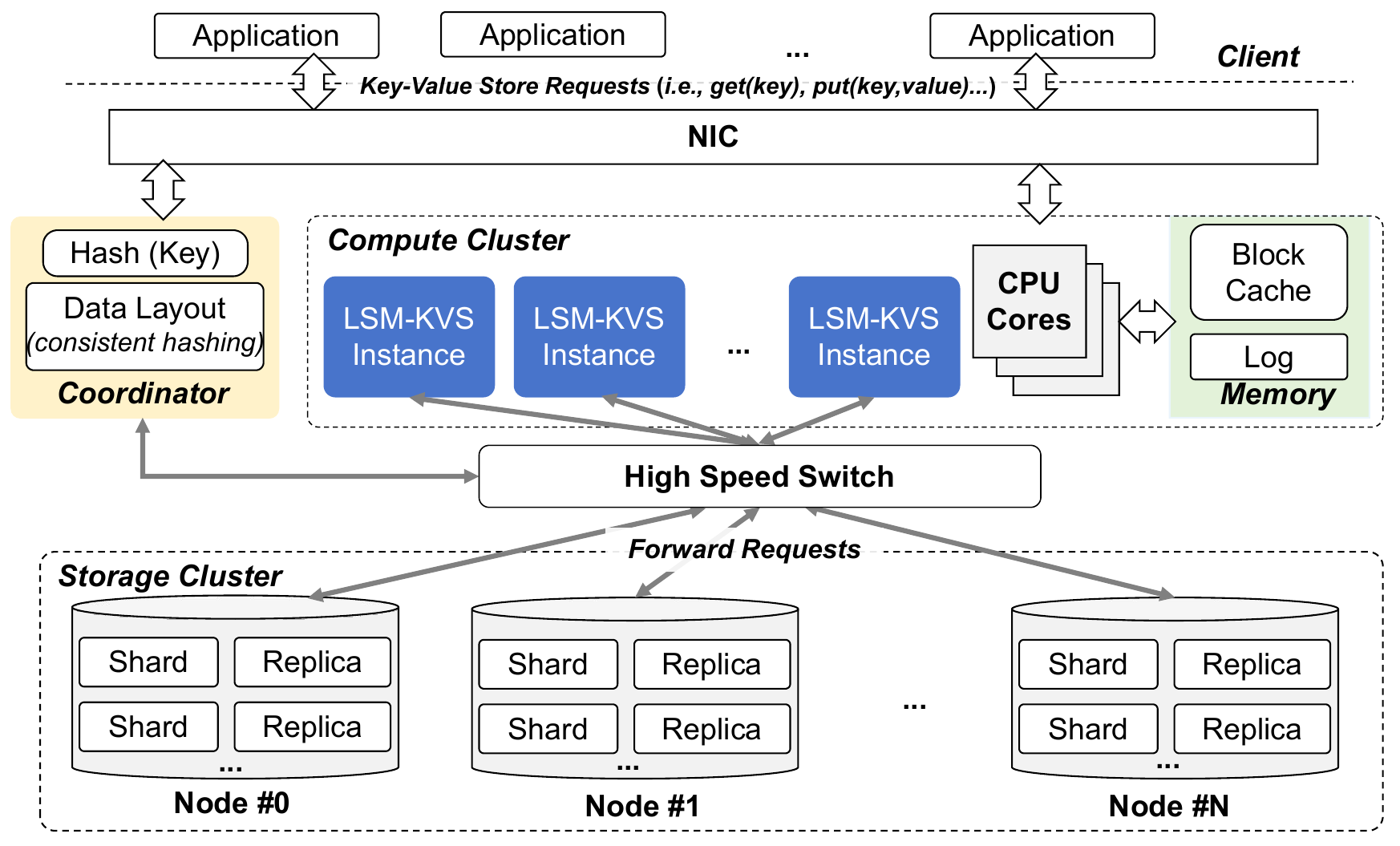}
\caption{Distributed LSM-KVS with Compute-Storage Disaggregation Architecture.}
\label{fig:LSM-KVS}
\end{figure*}

Within the distributed LSM-KVS designs, the compute-storage disaggregation architecture has emerged as a widely adopted solution in modern datacenters \cite{SIGMOD2024CaaS-LSM,FAST2024EBS,2023DisaggregatingRocksDB}.
As illustrated in Figure \ref{fig:LSM-KVS}, distributed LSM-KVS adopts a compute-storage disaggregation model, where multiple LSM-KVS instances in the compute cluster serve key-value requests from applications.
In this process, the coordinator is responsible for managing key-value store requests from applications in the client, distributing them across the storage cluster through high-speed switches, and maintaining data replication across the storage nodes.

This architecture decouples compute and storage resources into independent pools, enabling more flexible resource allocation and better operational efficiency.
The storage layer consists of specialized nodes that persistently maintain sharded data and handle local LSM-tree operations.
Each storage node in the disaggregated storage system is responsible for storing its assigned key ranges on local disk storage and for managing and querying its data.
This decentralized approach provides fault tolerance through data replication across nodes while enabling independent scaling of storage capacity and compute resources.
The compute layer focuses on request processing and coordination, offloading storage-intensive operations to storage nodes. 
Notably, compaction operations, traditionally a major CPU and I/O bottleneck, can be efficiently distributed across storage nodes. 
This separation enables compute nodes to continue responding to application requests while storage nodes handle background maintenance tasks.
Furthermore, this architecture makes maintenance and upgrades easier since nodes can be taken offline or updated without disrupting the system.
Additionally, compaction offloading in LSM-KVS to disaggregated storage can alleviate the pressure of local computation.
The compute nodes can focus on other operations by offloading the resource-intensive compaction process to the storage layer, improving overall system performance \cite{VLDB2015compaction,ToS2022ishbase}.

In this distributed setup, each storage node in the storage cluster maintains its LSM-tree structure with various levels ($L_0$-$L_n$), but the overall data distribution and level formation work differently than in single-node systems.
The storage cluster collectively forms a global LSM-tree where higher levels (typically $L_2$ and above) are distributed across nodes based on the sharding strategy. 
Lower levels ($L_0$-$L_1$) remain node-local for write performance, while upper levels are dynamically redistributed to maintain balanced data distribution. 
This hybrid approach combines the benefits of LSM-tree leveling with distributed scalability.
The physical manifestation of the storage cluster appears as a collection of individual LSM-trees, each managing a subset of the key space (\textit{shards}) \cite{OSDI2018sharding}.
During compaction, shards may be split or merged across nodes while maintaining level invariants --- each higher level covers an exponentially larger key range than the previous level, but now spanning multiple nodes.
The coordinator manages this distributed level formation through a global metadata service that tracks the location and boundaries of each shard at different levels.
For sharding strategies, distributed LSM stores typically employ either range-based or hash-based partitioning, each with distinct tradeoffs.
Specifically, range-based sharding groups contiguous key ranges together, preserving locality for range queries but potentially creating hotspots for popular key ranges. 
Hash-based sharding uniformly distributes keys across nodes using consistent hashing, providing better load balancing but sacrificing sequential access patterns.
Modern systems often implement hybrid approaches, such as dynamically adjusted range sharding with hash-based secondary distribution, or workload-aware sharding that adapts to access patterns.
Some advanced implementations also support tiered sharding, where different levels of the LSM-tree may use different sharding strategies optimized for their access characteristics.
For example, keeping lower levels hash-sharded for write distribution while upper levels use range sharding for read efficiency. 

This sharding strategy allows the system to scale horizontally by adding more nodes as data volume and workload demands increase.
Besides, it is linked with workload balancing mechanisms that monitor access patterns in real-time and automatically redistribute hot spots, ensuring no single node becomes a performance bottleneck. 
The distributed architecture simultaneously achieves two critical objectives: massive-scale data processing capacity through parallel execution across nodes, and robust fault tolerance via synchronous replication protocols that maintain multiple shard copies across failure domains.
Modern distributed LSM-KVS implementations employ a sophisticated combination of techniques to optimize performance at scale. 
Adaptive data compression algorithms dynamically adjust to balance CPU overhead with storage efficiency, while tiered storage hierarchies leverage the characteristics of different storage media to maximize cost-performance ratios.
The system coordinates global compaction scheduling to minimize interference with foreground operations and implements cross-node caching strategies that exploit access locality patterns. 

The distributed LSM-KVS architecture also faces several challenges in managing the LSM-tree, such as efficiently managing the compaction process across multiple nodes, minimizing network traffic, and ensuring data durability.
On the other hand, tens of thousands or even millions of tenants are in the data center, and these tenants have various behaviors and resource requirements \cite{SIGMOD2021LogStore}.
I/O isolation in multi-tenant scenarios becomes a critical issue for satisfying the performance requirements of each tenant.

\section{Review of LSM-KVS and Research Questions: New Challenges and Opportunities}\label{sec:review}
In this section, we review the design of existing LSM-KVS systems, answering the following five key research questions.
Table \ref{tab:questions} illustrates the details.
\begin{itemize}
    \item Compactions in LSM-tree: Where are we today?
    \item How to improve point lookup and range query performance?
    \item How can LSM-KVS adapt to emerging storage devices?
    \item How to cope with the evolution of LSM-KVS architectures: From single-node to distributed?
    \item How to ensure I/O isolation under multi-tenant scenarios?
\end{itemize}

%%%%%%%%%%%%%%%%%%%%%%%%%%%%%%%%%%%%%%%%%%%% ==Table Start== %%%%%%%%%%%%%%%%%%%%%%%%%%%%%%%%%%%%%%%%%%%
\begin{table*}[htbp]
\scriptsize
\caption{Research Questions and Limitations of Representative LSM-KVS Solutions\tnote{1}.\label{tab:questions}}
\vspace{-0.15in}
\begin{threeparttable}
\newcolumntype{L}{>{\raggedright\arraybackslash}X}
\newlength{\firstcolwidth}
\setlength{\firstcolwidth}{0.12\textwidth}
\begin{tabularx}{\textwidth}{p{\firstcolwidth}L}
\hline
%======================================= Question 1 =========================================
\rowcolor{c1}
\multicolumn{2}{l}{\textcolor{white}{\textbf{Question\#1: Compactions in LSM-tree: Where are we today?}}} 
\\\hline
\multirow{2}*{\textbf{Write-Stall}} & 
\textbf{Root Causes}: 
1. Unreasonable I/O resource allocation between key-value store operations and background compaction;
2. Suboptimal priority settings; 
3. Compaction strategy limitations. \\
& \textbf{Solutions}:
\begin{itemize}[leftmargin=*,noitemsep,topsep=0pt]
\item \textit{Write-throttling/Rate-limiter}: SILK\cite{ATC2019SILK} (ATC'19), Vigil-KV\cite{ATC2022vigilkv} (ATC'22), ADOC\cite{FAST2023ADOC} (FAST'23)
\item \textit{Compaction speed/Pipeline compaction}: Zhang et al.\cite{IPDPS2014pipelinecompaction} (IPDPS'14), gLSM\cite{ToS2024gLSM} (ToS'24)
\item \textit{Resource-aware I/O scheduling}: SILK\cite{ATC2019SILK} (ATC'19), SpanDB\cite{FAST2021SpanDB} (FAST'21), Vigil-KV\cite{ATC2022vigilkv} (ATC'22), ADOC\cite{FAST2023ADOC} (FAST'23), STEM\cite{ICDE2024STEM} (ICDE'24)
\item \textit{Selective/Delayed/Partial compaction}: dCompaction\cite{pan2017dcompaction} (JCST'17), TRIAD\cite{ATC2017TRIAD} (ATC'17), Dostoevsky\cite{SIGMOD2018dostoevsky} (SIGMOD'18), NVLSM\cite{ToS2021NVLSM} (ToS'21), BlockDB\cite{ICDE2022BlockDB} (ICDE'22), Spooky\cite{VLDB2022Spooky} (VLDB'22)
\item \textit{Compaction granularity}: DOPA-DB\cite{HotStorage2024advocating} (HotStorage'24)
\item \textit{Data layout optimization}: PebblesDB\cite{SOSP2017pebblesdb} (SOSP'17), vLSM\cite{2024vLSM}, SA-LSM\cite{VLDB2022SALSM} (VLDB'22), Saxena et al.\cite{ICDE2023RealTime} (ICDE'23)
\item \textit{Hardware offloading}: Zhang et al.\cite{FAST2020FPGA} (FAST'20), Sun et al.\cite{ICDE2020FPGA} (ICDE'20), FaaS Compaction\cite{CLUSTER2021Supporting} (CLUSTER'21), D2Comp\cite{TACO2024D2Comp} (TACO'24), PStore\cite{TECS2024PStore} (TECS'24), gLSM\cite{ToS2024gLSM} (ToS'24), Edgepilot\cite{CLOUD2024Coordinating} (CLOUD'24)
\item \textit{Heterogeneous storage}: SplitKV\cite{HotStorage2020SplitKV} (HotStorage'20), SpanDB\cite{FAST2021SpanDB} (FAST'21), Prism\cite{ASPLOS2023Prism} (ASPLOS'23), PrismDB\cite{ASPLOS2023PrismDB} (ASPLOS'23), MirrorKV\cite{SIGMOD2023mirrorkv} (SIGMOD'23), Chen et al.\cite{ICDE2023Workload} (ICDE'23)
\vspace{-0.1in}
\end{itemize}
\\\midrule
\textbf{Unpredictable Latency} &
\textbf{Root Causes}:
1. Complex parameter tuning (hundreds of parameters); 
2. Inability to adapt to dynamic workload characteristics. 
\\
& \textbf{Solutions}:
\begin{itemize}[leftmargin=*,noitemsep,topsep=0pt]
\item \textit{Auto-tuning frameworks}: Endure\cite{VLDB2022Endure} (VLDB'22), Dremel\cite{2022Dremel}, ADOC\cite{FAST2023ADOC} (FAST'23), ELMo-Tune\cite{HotStorage2024ELMo-Tune,ELMo-Tune-V2} (HotStorage'24)
\item \textit{Workload-aware designs}: TRIAD\cite{ATC2017TRIAD} (ATC'17), Mutant\cite{SoCC2018Mutant} (SoCC'18), Sharding\cite{OSDI2018sharding} (OSDI'18), Cao et al.\cite{FAST2020RocksDBFacebook} (FAST'20), CruiseDB\cite{ICDE2021CruiseDB} (ICDE'21), Endure\cite{VLDB2022Endure} (VLDB'22), Cerberus\cite{TECS2023multitenant} (TECS'23), Calcspar\cite{ATC2023calcspar} (ATC'23), ELECT\cite{FAST2024ELECT} (FAST'24), Moose\cite{SIGMOD2024Moose} (SIGMOD'24), ELMo-Tune\cite{HotStorage2024ELMo-Tune,ELMo-Tune-V2} (HotStorage'24), RusKey\cite{SIGMOD2023learning} (SIGMOD'23)
\item \textit{Tailored LSM-KVS designs}: VT-tree\cite{FAST2013VTtree} (FAST'13), LSM-trie\cite{ATC2015LSM-trie} (ATC'15), SlimDB\cite{VLDB2017SlimDB} (VLDB'17), Sharding\cite{OSDI2018sharding}, DiffKV\cite{ATC2021DiffKV} (ATC'21), LogStore\cite{SIGMOD2021LogStore} (SIGMOD'21), TWEEZER\cite{FAST2022TWEEZER} (FAST'22)
\vspace{-0.1in}
\end{itemize} 
\\\midrule
\textbf{Write} & 
\textbf{Root Causes}: 
1. Out-of-place update mechanism; 
2. Delayed actual data deletion until compaction. \\
\textbf{Amplification} & \textbf{Solutions}:
\begin{itemize}[leftmargin=*,noitemsep,topsep=0pt]
\item \textit{Write amplification reduction}: NoFTL-KV\cite{EDBT2018noftlkv} (EDBT'18), MatrixKV\cite{ATC2020MatrixKV} (ATC'20), TriangleKV\cite{TPDS2022TriangleKV} (TPDS'22), BlockDB\cite{ICDE2022BlockDB} (ICDE'22)
\item \textit{Privacy-aware deletion}: Lethe\cite{SIGMOD2020Lethe} (SIGMOD'20)
\vspace{-0.05in}
\end{itemize} 
\\\hline

%======================================= Question 2 =========================================
\rowcolor{c1}
\multicolumn{2}{l}{\textcolor{white}{\textbf{Question\#2: How to improve point lookup and range query performance?}}} \\
\hline
\textbf{Read Amplification}
& 
\textbf{Root Causes}: 
1. Need to read multiple SSTables for key lookup; 
2. Bloom filter limitations; 
3. Read-write performance trade-offs. \\
& \textbf{Solutions}:
\begin{itemize}[leftmargin=*,noitemsep,topsep=0pt]
\item \textit{Point lookup optimization}: bLSM\cite{SIGMOD2012bLSM} (SIGMOD’12), Monkey\cite{SIGMOD2017Monkey} (SIGMOD'17), TRIAD\cite{ATC2017TRIAD} (ATC'17), ElasticBF\cite{ATC2019ElasticBF} (ATC'19), Bourbon\cite{OSDI2020from} (OSDI'20), Zhu et al.\cite{DAMON2021Reducing} (DAMON'21), Moose\cite{SIGMOD2024Moose} (SIGMOD'24)
\item \textit{Range query optimization}: SuRF\cite{SIGMOD2018SuRF} (ICMD’18), Rosetta\cite{SIGMOD2020Rosetta} (SIGMOD'20), REMIX\cite{FAST2021REMIX} (FAST'21), Disco\cite{SIGMOD2025Disco} (SIGMOD'25), GRF\cite{SIGMOD2024GRF} (SIGMOD'24)
\item \textit{Secondary index optimization}: Lethe\cite{SIGMOD2020Lethe} (SIGMOD'20), SineKV\cite{ICDCS2020SineKV} (ICDCS'20), Wang et al.\cite{ATC2023Revisiting} (ATC'23), Perseid\cite{ToS2024Perseid} (ToS'24)
\item \textit{Cache optimization}: LSbM-tree\cite{ICDCS2017LSbM-tree} (ICDCS'17), AC-Key\cite{ATC2020ACKey} (ATC'20), Leaper\cite{VLDB2020Leaper} (VLDB'20), Calcspar\cite{ATC2023calcspar} (ATC'23)
\vspace{-0.05in}
\end{itemize} 
\\\hline

%===================================== Question 3 ===========================================
\rowcolor{c1}
\multicolumn{2}{l}{\textcolor{white}{\textbf{Question\#3: How can LSM-KVS adapt to emerging storage devices?}}} \\
\hline
\textbf{Hardware Utilization} & 
\textbf{Root Causes}:
1. Underutilized parallelism in NVMe SSDs;
2. Need for ZNS storage adaptation;
3. Persistent memory integration challenges. \\
& \textbf{Solutions}:
\begin{itemize}[leftmargin=*,noitemsep,topsep=0pt]
\item \textit{Hardware efficiency}: LOCS\cite{EuroSys2014LOCS} (EuroSys'14), BlueCache\cite{VLDB2016BlueCache} (VLDB'16), NoFTL-KV\cite{EDBT2018noftlkv} (EDBT'18), Check-In\cite{ISCA2020CheckIn} (ISCA'20), KVIMR\cite{ATC2021KVIMR} (ATC'21), Dotori\cite{VLDB2023Dotori} (VLDB'23)
\item \textit{Zoned namespace storage}: LSM\_ZGC\cite{HotStorage2020LSM_ZGC} (HotStorage'20), Jung et al.\cite{HotStorage2022LLCompaction} (HotStorage'22), SplitZNS\cite{TACO2023SplitZNS} (TACO'23), WALTZ\cite{VLDB2023waltz} (VLDB'23)
\item \textit{Persistent memory}: NoveLSM\cite{ATC2018NoveLSM} (ATC'18), Wisckey\cite{OSDI2020from} (OSDI'20), MatrixKV\cite{ATC2020MatrixKV} (ATC'20), Clover\cite{ATC2020Clover} (ATC'20), L2SM\cite{ICDE2021L2SM} (ICDE'21), Kreon\cite{ToS2021Kreon} (ToS'21), NVLSM\cite{ToS2021NVLSM} (ToS'21), ChameleonDB\cite{EuroSys2021ChameleonDB} (EuroSys'21), Pacman\cite{ATC2022Pacman} (ATC'22), CacheKV\cite{ICDE2023CacheKV} (ICDE'23), ZigZagDB\cite{ICDE2024ZigZagDB} (ICDE'24)
\vspace{-0.05in}
\end{itemize} 
\\\hline

%==================================== Question 4 ============================================
\rowcolor{c1}
\multicolumn{2}{l}{\textcolor{white}{\textbf{Question\#4: How to cope with the evolution of LSM-KVS architectures: From single-node to distributed?}}} \\
\hline
\textbf{Resource} & 
\textbf{Root Causes}:
1. High network overhead in distributed systems;
2. Lack of global resource management. \\
\textbf{Underutilization} & \textbf{Solutions}:
\begin{itemize}[leftmargin=*,noitemsep,topsep=0pt]
\item \textit{Distributed KVS designs}: Hailstorm\cite{ASPLOS2020Hailstorm} (ASPLOS'20), Nova-LSM\cite{SIGMOD2021NovaLSM} (SIGMOD'21), RocksMash\cite{TCAO2022building} (TACO'22), DEPART\cite{FAST2022DEPART} (FAST'22), IS-HBase\cite{ToS2022ishbase} (ToS'22), LightPool\cite{HPCA2024LightPool} (HPCA'24), Caas-LSM\cite{SIGMOD2024CaaS-LSM} (SIGMOD'24)
\vspace{-0.05in}
\end{itemize} \\\hline

%===================================== Question 5 ===========================================
\rowcolor{c1}
\multicolumn{2}{l}{\textcolor{white}{\textbf{Question\#5: How to ensure I/O isolation under multi-tenant scenarios?}}} \\
\hline
\textbf{Multi-tenant} & 
\textbf{Root Causes}: 
1. Lack of I/O path isolation between tenants. \\
\textbf{Conflict} & \textbf{Solutions}:
\begin{itemize}[leftmargin=*,noitemsep,topsep=0pt]
\item \textit{Performance isolation}: Iso-KVSSD\cite{HotStorage2021Iso-KVSSD} (HotStorage'21), OceanBase\cite{VLDB2022oceanbase,VLDB2023oceanbase} (VLDB'22)
\item \textit{Efficient sharing}: Jaliminche et al.\cite{SoCC2023enabling} (SoCC'23)
\vspace{-0.1in}
\end{itemize} \\\bottomrule
\end{tabularx}

\begin{tablenotes}
\item[1] Most reviewed works were published between 2020-2025, covering top-tier conferences and journals in the field.
\end{tablenotes}
\end{threeparttable}
\end{table*}

The conventional LSM-tree design leverages sequential writes to achieve high write throughput, a critical optimization for hard disk drives (HDDs) where random writes perform significantly worse than sequential operations.
However, this design can degrade read performance due to the need to search multiple LSM-tree levels.
Existing studies have primarily focused on optimizing LSM-tree compaction strategies, either by improving I/O resource allocation, tuning the configurable parameters, or refining compaction strategies themselves.
However, compactions are still everywhere, posing a potential risk of decreasing user access performance, particularly under burst workload scenarios or in resource-constrained environments.

Modern LSM-KVSs are increasingly deployed in distributed environments, such as compute-storage separation or within disaggregated storage clusters.
In these settings, the factors that influence performance are not confined to allocating I/O resources at a local level.
They also include the allocation of available resources across different nodes and even across data center nodes, the management of network traffic, and the accommodation of multiple tenants, each with its own diverse and dynamic characteristics.
Moreover, with the proliferation of AI applications, the volume of datasets has increased exponentially.
Concurrently, the computational demands of users have also increased significantly.
These applications often require the processing of large-scale data and high computation requirements for users, which places additional challenges on the LSM-KVS.
By addressing these challenges, the potential of LSM-trees can be unlocked in distributed, AI-driven systems, ensuring that they remain a robust and efficient solution for data storage and data retrieval.

In the following, we review and discuss LSM-tree designs, including the compaction process, key-value store operations like query and delete, and a new perspective on modern LSM-KVS designs from distributed LSM-KVS, hardware-assisted LSM-KVS, and heterogeneous storage devices.

\subsection{Compactions in LSM-trees: Where Are We Today?} 
Compaction operations, which ensure data sequentiality and space collection in the LSM-tree, have historically been a complex and persistent issue that has yet to be solved in the LSM-KVS.
In the following, we first review compaction optimization \cite{VLDB2021constructing,SIGMOD2023SplinterDB,SIGMOD2025RethinkingCompaction,DATE2025RemapCom} from the traditional perspective, including 1) \textit{how to allocate the resources between foreground and internal operations?} 2) \textit{how to obtain the suitable parameter configuration of LSM-tree?} 3) \textit{how to make efficient scheduling by setting suitable priorities for host and internal operations?} and 4) \textit{how to design a performance-oriented or space-oriented compaction strategy?}
Then, we discuss the optimization of key-value store operations, including the point lookup, range query operations, and delete operations.
Here, we should consider \textit{how to achieve fast point lookup and range query?} and \textit{how to trade off the read and write performance based on the LSM-tree management?}
On the other hand, we consider new challenges and remaining issues of state-of-the-art studies from the perspectives of emerging hardware and architectures as well as various applications to analyze the reasons behind LSM-KVS designs, including 1) \textit{how to achieve high resource utilization in the distributed architectures?} 2) \textit{how to leverage the features and make efficient use of emerging hardware?} 3) \textit{how to ensure resource isolation and sharing among multiple tenants?} 4) \textit{how to achieve dynamic LSM-tree designs based on the workload?}

\subsubsection{Traditional Perspective}
The primary reason for the impact of compaction on LSM-trees is that host read and write operations can temporarily be hindered during the compaction process.
This occurs as the key-value store performs the complex tasks of reorganizing data across different levels of the LSM-tree structure.
While such reorganization is essential for preserving data sequentiality and space efficiency, it can cause a temporary performance drop as system resources are redirected to handle background compaction tasks.
To mitigate the performance impact of compaction processes, several optimization techniques can be employed as follows:

\textbf{Resource Allocation.}
The compaction-induced write-stall issue is a resource contention problem.
Therefore, how to reasonably allocate resources between key-value store operations and background flush and compaction operations has become a solution to alleviate the impact on system performance.
One simple solution is \textit{write throttling}, which limits the resources allocated to host write operations to prevent them from overwhelming the processing of internal operations in key-value stores \cite{RocksDB,ATC2019SILK}.
Write throttling occurs when the internal flushing and compaction process cannot keep up with the host writes.
In extreme cases, it will completely stall writing.
Take RocksDB as an example.
There are two thresholds (slowdown writes threshold and stop writes threshold) to control the write rates.
If the number of SSTables at $L_0$ reaches the slowdown writes threshold, the host writes begin to stall; if it reaches the stop writes threshold, the host writes are entirely stalled. 
Additionally, we can utilize the pending compaction bytes to indicate a write stall.
If the pending compaction bytes reach the soft threshold, host writes will stall; if they exceed the hard threshold, host writes will be completely halted.
In addition, another similar solution is \textit{rate limiter} \cite{RocksDB}, which proactively controls the rate of incoming writes to prevent the system from being overloaded by burst workloads and maintain stable performance.
Furthermore, SILK \cite{ATC2019SILK,TOCS2020silk+} proposed to dynamically allocate bandwidth between client and internal operations to prevent latency spikes during the background compaction process.
However, SILK is designed for write-intensive workloads and may have less performance impact on read- or scan-intensive workloads.
Furthermore, to consider both the read and write performance, Vigil-KV \cite{ATC2022vigilkv} proposed to manage write rates in the NVMe SSD (solid-state drive) by implementing a predictable latency mode interface that pauses all write requests during the deterministic time window to ensure consistent read performance.
However, the above write throttling or rate limiter methods not only impact host performance but also introduce a risk of accumulated writes, increasing the frequency of flushing and compaction operations in the next period.

Some prior works proposed to allocate resources between key-value store operations and background tasks based on current workloads and available background resources, allowing the system to handle high write volumes by increasing the number of threads for compaction, thereby improving overall performance.
Furthermore, some systems employ adaptive algorithms that continuously monitor performance metrics and dynamically modify the thread count allocated to background tasks to ensure maximum throughput and prevent sudden increases in latency.
For example, ADOC \cite{FAST2023ADOC} proposed to avoid write-stall issues by varying two parameters (i.e., thread count and batch size) to allocate the thread count based on the dataflow.
These demonstrate how resource allocation can effectively address the I/O contention between foreground user operations and background operations in LSM-tree-based storage systems.

The design of LSM-tree compaction will increasingly focus on dynamic, fine-grained resource allocation to balance performance and efficiency.
Future systems could leverage multi-dimensional resource partitioning that dynamically distributes CPU threads, I/O bandwidth, and memory bandwidth between foreground operations and background compaction based on real-time workload demands.
Hardware accelerators could enable dedicated resource pools for compaction, isolating their impact on foreground workloads.
The machine learning-driven resource manager predicts the optimal allocation ratio and automatically adjusts parameters such as the number of compaction threads and I/O allocation.
Workload-aware resource allocation protocols can dynamically adjust the allocation between writes, reads, and compaction based on SLA requirements.
The future direction of LSM-trees in resource allocation is not as a static configuration but as a continuously optimized parameter, enabling the system to automatically find the most efficient trade-off between write amplification, read performance, and space utilization under varying workloads.

\textbf{I/O Scheduling.}
Apart from allocating resources to balance the internal management of LSM-trees and host I/O operations, setting the operation priority through I/O scheduling is also an efficient way to mitigate the impact of background operations, enhance compaction throughput by ensuring sufficient I/O bandwidth, and distribute the workload to prevent resource contention.
The I/O scheduler prioritizes I/O operations in a way that supports these internal compaction rules efficiently, without causing performance bottlenecks.
This is challenging because the I/O scheduler must be deeply integrated with the LSM-tree's internal mechanisms, requiring precise awareness of multiple factors: the current compaction levels, the varying sizes of SSTables (which impact the compaction granularity), and the intricate dependencies between concurrent compaction tasks.
It is difficult to make optimal scheduling decisions that enhance compaction performance without negatively impacting other key-value store operations.

Several prior works proposed to prevent latency spikes from compaction operations by optimizing I/O scheduling processes.
For example, Zhang et al. \cite{IPDPS2014pipelinecompaction} proposed pipelining the compaction operations so that the internal compaction process can be executed in parallel and completed quickly.
SILK \cite{ATC2019SILK,TOCS2020silk+} resolved resource contention among host writes, flushes, and compactions by prioritizing critical internal operations and allowing for the preemption of less critical tasks to prevent latency spikes.
Furthermore, STEM \cite{ICDE2024STEM} presented a streamlined algorithm designed for efficient, large-scale data compaction.
STEM employs a multi-unit pipeline and dynamic scheduling to significantly enhance compaction throughput and reduce write stalls, thereby improving overall write performance.
Recent work also leverages the properties of the NVMe protocol (e.g., I/O determinism) to enable the processing of different operations within a specific window, thereby mitigating performance impact.
For example, Vigil-KV \cite{ATC2022vigilkv} proposed to dynamically schedule internal tasks, such as flush and compaction, during non-deterministic time windows to minimize their interference with client read requests. This ensures that these internal operations are executed efficiently while maintaining optimal read performance.
In addition, SpanDB \cite{FAST2021SpanDB} proposed to adapt RocksDB to utilize high-speed SSDs selectively, relocating write-ahead logs and top LSM-tree levels to a smaller, faster NVMe SSD for improved performance.
As we can see, LSM-KVS optimization via I/O scheduling strategically distributes tasks across storage devices or schedules them temporally. This approach effectively mitigates resource contention between host I/O operations and background compactions.

\textbf{Auto-Tuning.}
Modern LSM-KVS offers extensive configurability, with tunable parameters spanning compaction strategies, data layouts, MemTable sizing, and more \cite{VLDBJ2024towards}. 
However, achieving optimal performance through parameter tuning is not an easy problem for the following reasons.
First, the parameter dataset is large in modern large-scale LSM-KVS, with over 100 parameters \cite{HotStorage2024ELMo-Tune,ELMo-Tune-V2,ICDCS2020kill}.
Second, various factors should be considered, including complex and dynamic workloads, hardware, multi-tenants with different requirements, and emerging different architectures.
Third, the application, underlying hardware, and architecture are frequently updated, especially in cloud environments with bursty workloads.

Traditional approaches relying on manual tuning or limited parameter optimization are both constrained in performance and labor-intensive. 
Some prior works consider the workload characteristic and construct models to guide the parameter tuning.
For example, Endure \cite{VLDB2022Endure,VLDBJ2024towards} proposed to use dynamic tuning based on workload variations, addressing performance degradation due to workload uncertainty by maximizing worst-case throughput across a range of possible workloads.
Dremel \cite{2022Dremel} addressed the challenge of configuring RocksDB by using a Multi-Armed Bandit model to adaptively and quickly find optimal settings, and employs fused features to manage the vast parameter space, an online bandit model to adjust to varying workloads and hardware, and multi-fidelity evaluation with upper-confidence-bound sampling to expedite the configuration process.
Some works conclude preliminary experiments and perform parameter tuning based on these conclusions.
For example, ADOC \cite{FAST2023ADOC} proposed an automatic tuning framework that adjusts system configurations, such as thread count and batch size, to minimize data overflow, achieving significant reductions in write stall duration and substantial performance improvements compared to auto-tuned and manually optimized systems.
Some related works proposed using machine-learning methods or large language models (LLMs) to predict parameter values and realize parameter auto-tuning.
For example, ELMo-Tune \cite{HotStorage2024ELMo-Tune,ELMo-Tune-V2} proposed to leverage the LLM for parameter prediction.
While AI and LLMs are the current hot topics and main trends in applications, applying these methods to LSM-trees enables the optimal parameter setting based on iterations. However, this approach also incurs additional tuning time costs and storage overhead for generating the best options from LLMs and storing collected datasets for prompts.
We constructed an experiment to evaluate the tuning time cost based on ELMo-Tune, revealing that the tuning process requires approximately one hundred seconds when performed by DeepSeek V3 \cite{DeepSeek} under db\_bench evaluation. 
This duration makes the method unsuitable for deployment in real-time systems.

\textbf{Compaction Optimization.} 
Existing studies have focused on optimizing compactions of LSM-KVS in four aspects \cite{VLDB2021constructing,SIGMOD2023SplinterDB}: when to trigger, which data to compact, compaction granularity, and data layout.

\textit{1) When to Trigger:}
The timing of compaction operations in LSM-KVS systems is a critical parameter that significantly influences system performance.
Compaction, which merges data from multiple SSTables and removes obsolete entries, is typically triggered based on specific thresholds, such as when a certain number of SSTables accumulate or when a percentage of the data is marked for deletion \cite{SIGMOD2020Lethe}.
Compaction triggers at suitable times are essential to minimize their impact on read and write operations.
Ideally, it is performed during off-peak hours or when system load is lower to reduce contention for I/O resources.
However, compactions are often triggered passively, especially in large-scale data centers with burst write scenarios. 
Some studies have proposed using selective or delayed compaction to optimize the performance of LSM-tree-based systems.
For example, dCompaction \cite{pan2017dcompaction} proposed a new compaction scheme that decreases write amplification by postponing some compactions and gathering them into subsequent compactions.
Dostoevsky \cite{SIGMOD2018dostoevsky} proposed to optimize space-time trade-offs in LSM-tree designs by adaptively removing unnecessary merging operations, improving performance and storage efficiency.
TRIAD \cite{ATC2017TRIAD} proposed to optimize compaction by deferring and batching compaction operations until the overlap between files becomes large enough to merge a large number of duplicate keys, thereby reducing the frequency and intensity of compaction operations.
Most LSM-KVSs use the manual parameter setup to decide when to trigger flushing or compaction operations.
However, in modern scenarios, the workload is dynamically changed, and performance and amplification should also be traded off based on the workload characteristics.
Therefore, it becomes more difficult to effectively set the timing of the compaction trigger.

\textit{2) Which Data:}
When executing compaction operations, one important question is which data to select for each compaction.
This is decided mainly by the overlapped SSTables based on the victim SSTable.
When we select an SSTable for compaction, we should compare the overlap ratio of each SSTable at the current level with that of all SSTables in the next level.
The SSTable with the highest overlap ratio is often chosen as the victim SSTable for compaction, as it is likely to contain the most obsolete data that needs to be collected.
This approach ensures that the compaction process focuses on SSTables, yielding the greatest benefits in data deduplication and storage space release.
Moreover, some works further refine the selection of SSTables during compactions by considering additional factors such as the age of the data, the frequency of data access, and the overall system performance impact.
For example, Spooky \cite{VLDB2022Spooky} solved the space amplification under global compaction and the write amplification under partial compaction, and further considers the lifetime of files to combine the same lifetime of the file.
BlockDB \cite{ICDE2022BlockDB} proposed to use block compaction to address the issues of write amplification and block-cache invalidation associated with conventional table compaction.

\textit{3) Compaction Granularity:}
Compaction granularity refers to the amount of data processed in a single compaction operation within an LSM-tree, playing a critical role in balancing write amplification (i.e., the amount of data written to disk relative to the original data size) and performance impact in LSM-trees.
Smaller granularity reduces write amplification but may increase compaction frequency, while larger granularity can cause significant I/O and CPU overhead, affecting concurrent operations.
Recent work has proposed dynamically adjusting compaction granularity based on workload patterns \cite{HotStorage2024advocating}.
For example, DOPA-DB \cite{HotStorage2024advocating} proposed to dynamically adapt compaction strategies by varying compaction sizes based on detected workload patterns, using a compaction size recommender to minimize write stalls and read latency.
This allows for more efficient LSM-tree management, aligning compaction operations with the specific demands of the workload.
However, implementing dynamic compaction granularity introduces complex algorithms to accurately analyze workload patterns and perform timely adjustments.
Additionally, the system should allocate sufficient computational resources for monitoring and decision-making without significantly impacting overall system performance.
It is also crucial to ensure the stability and predictability of the system while making real-time changes to compaction granularity.
Despite these challenges, the potential benefits of improved performance and reduced write amplification make dynamic compaction granularity a valuable area of research in LSM-tree optimizations.

\textit{4) Data Layout:}
For LSM-tree designs, leveling and tiering are two distinct data layouts to manage data organization and compaction.
For the leveling data layout, data is organized into different levels, and compaction operations are performed across these levels. 
Each level contains sorted runs of data, and compaction in $L_i$ involves merging files that are marked for compaction within that level and then promoting the merged result to $L_{i+1}$. 
Leveling ensures strong orderliness.
Apart from $L_0$, each level only includes one sorted run.
Leveling is crucial for managing write amplification and optimizing read performance.
For the tiering data layout, each level in the LSM-tree can contain multiple sorted runs, which may have overlapping key ranges.
When compaction is triggered in $L_i$, all sorted runs within that level are merged.
The outcome of this compaction is a new sorted run that is written to $L_{i+1}$, without disturbing the existing sorted runs in $L_{i+1}$. 
In real scenarios, $L_0$ often adopts a tiering layout while other layers use a leveling layout (e.g., RocksDB).
To optimize data layouts, PebblesDB \cite{SOSP2017pebblesdb} addresses the high write amplification problem in key-value stores by introducing a novel data structure that reduces data rewrites and increases write throughput.
SA-LSM \cite{VLDB2022SALSM} applied survival analysis, a statistical method typically used in biostatistics, to optimize data layout in LSM-KVS for managing cold data.
Saxena et al. \cite{ICDE2023RealTime} introduce the concept of a real-time LSM-tree, which leverages the natural lifecycle-aware characteristics of LSM-trees to support hybrid data layouts in analytics systems.
It proposes a modified LSM-tree structure where different levels have varying data layouts, from row-oriented for recent data to column-oriented for older data.
vLSM \cite{2024vLSM} studies the impact on tail latency by modifying the width and length of the compaction chain.
Among them, the width is adjusted using small SSTables, eliminating the tiering compaction in $L_0$, while the length is adjusted by a larger growing factor between $L_1$ and $L_2$.

In conclusion, compaction is a potential internal operation of LSM-tree architectures, triggering decisions by various factors, such as workload characteristics, underlying LSM-tree designs, hardware features, and the impact of other applications.
As the dataset scales and more storage nodes are required to service key-value store operations, the challenges remain, and the LSM-tree optimization becomes difficult.
This is because 1) the workload characteristics among tens of thousands or even millions of tenants should be identified to provide online and adaptive LSM-tree designs; 2) the update and iteration of the underlying new hardware provide new ideas for LSM-tree design; 3) the distributed architecture involves more impact factors and opportunities.
In the next subsection, we will further discuss the LSM-KVS from new perspectives.

\subsubsection{New Perspective}
When LSM-tree is applied to large-scale distributed key-value stores and multi-tenant environments, the complexity of managing data across multiple storage nodes and tenants increases the difficulty of compactions.
First, the resources among different nodes are underutilized because the applications in each storage node are significantly different, and dynamic workloads are common in data centers.
Second, the multi-tenant characteristics are not fully explored.
On the one hand, the isolated LSM-tree management is used to construct a resource-sharing architecture.
On the other hand, the most common solution to avoid resource contention among multiple tenants is resource isolation, with the isolated LSM-tree management allocated to each tenant.
Third, network traffic in large-scale distributed scenarios should also be considered to execute efficient compaction operations.
When the underlying storage utilizes multiple storage media, leveraging the storage characteristics to implement an LSM-tree data layout has also become a key focus of existing work.
In the following, we will list the state-of-the-art compaction solutions for deploying LSM-trees in distributed architecture, emerging hardware, or heterogeneous storage systems.

\textbf{Large-Scale Distributed Key-Value Stores.}
For large-scale distributed key-value stores, we can leverage the advantages of local and remote/cloud storage to improve system performance.
Local storage can provide rapid access to data with its fast read and write speeds, making it ideal for applications that require immediate data processing.
However, this speed comes at a price, as local storage solutions are often more expensive and have relatively limited capacity compared to remote cloud storage options.
On the other hand, remote cloud storage can offer a cost-efficient solution with expansive storage capacities, suitable for large-scale data archiving and backup.
While it excels in cost and space, its data retrieval speeds are generally slower due to the latency of data transfer over networks, making it less suitable for time-sensitive applications that demand quick data access.

Recent work has optimized compaction in LSM-KVS for a distributed storage infrastructure.
For example, Dong et al. \cite{FAST2021Evolution} discussed the evolution of development priorities in RocksDB, highlighting how it has adapted to optimize for write amplification, space utilization, and CPU efficiency in response to hardware trends and real-world application demands, ultimately improving performance for large-scale distributed systems.
Hailstorm \cite{ASPLOS2020Hailstorm} solved the challenges of throughput and latency in distributed LSM-based databases by leveraging a distributed filesystem that separates storage from processing, thereby enabling storage pooling and offloading of compaction tasks to remote nodes. 
This approach not only balances the workload across instances without resharding but also enhances system throughput and latency by distributing the impact of background tasks.
Furthermore, Nova-LSM \cite{SIGMOD2021NovaLSM} proposed a disaggregated LSM-KVS that leverages RDMA communication to decouple storage from processing, enhancing scalability and performance through dynamic compaction and load balancing. It significantly outperforms monolithic data stores, such as LevelDB \cite{LevelDB} and RocksDB \cite{RocksDB}, especially under skewed access patterns.
DEPART \cite{FAST2022DEPART} proposed a distributed key-value store that implements replica decoupling to optimize storage management and I/O costs by using a two-layer log, achieving significant performance improvements over Cassandra across various operations and consistency levels.
On the other hand, specialized optimizations include TWEEZER's \cite{FAST2022TWEEZER} memory protection improvements for confidential computing environments and IS-HBase's \cite{ToS2022ishbase} network traffic reduction in disaggregated infrastructures via I/O offloading.
RocksMash \cite{TCAO2022building} proposed hybrid local-cloud storage tiering to balance performance and cost-efficiency, while Caas-LSM \cite{SIGMOD2024CaaS-LSM} proposed to decouple compaction into stateless services with adaptive control planes.
Most recently, LightPool \cite{HPCA2024LightPool} is a storage pool architecture designed to improve storage resource utilization in cloud-native distributed key-value stores like OceanBase \cite{VLDB2022oceanbase,VLDB2023oceanbase}.
LightPool aggregates cluster storage into a unified pool, managed centrally, using NVMe-oF for high-performance sharing and integrating with Kubernetes for flexible resource allocation.
This design addresses the issues of low storage utilization and performance bottlenecks associated with disaggregated storage solutions.
Disaggregated memory systems face challenges with ordered key-value stores due to network overhead and limited compute resources in memory nodes \cite{FAST2023ROLEX}.

Distributed key-value stores have developed to address challenges in scalability, performance, and resource efficiency.
Recent research has shown that novel architectures can effectively combine the advantages of local and remote storage by leveraging techniques such as storage pooling, hardware offload compaction, and resource disaggregation.
However, as distributed key-value stores continue to scale to meet the needs of cloud-native applications and global services, several key challenges remain.
These challenges include maintaining low-latency access between geographically distributed nodes, ensuring strong consistency without sacrificing availability, and optimizing energy efficiency in increasingly heterogeneous hardware environments.

\textbf{In-Storage Compaction.}
Many existing works \cite{FAST2020FPGA,ICDE2020FPGA,TACO2024D2Comp,TECS2024PStore,ToS2024gLSM,SIGMOD2024CaaS-LSM} propose using additional computational components, such as data processing units (DPUs), general-purpose graphics processing units (GPGPUs), and field-programmable gate arrays (FPGAs), to offload compaction tasks from the CPU and alleviate the CPU's resource contention, thereby enhancing system performance.
Among these computational components, each leverages its unique advantages in various scenarios.
DPUs specialize in data-intensive operations, offering high throughput and low latency, which benefits the large data volumes processed during compaction. 
GPGPUs are applied to parallel processing, making them effective for tasks that can be parallelized, such as sorting and merging datasets.
FPGAs provide the flexibility and customization needed for specialized compaction algorithms.
For example, Zhang et al. \cite{FAST2020FPGA} achieved significant compaction speedups for short key-value pairs through FPGA offloading, and Sun et al. \cite{ICDE2020FPGA} developed an FPGA compaction engine using key-value separation and index-data block separation strategies to maximize bandwidth utilization.
DPU-based approaches have also proven effective, with D2Comp \cite{TACO2024D2Comp} demonstrating substantial performance improvements through fine-grained dynamic compaction offloading in RocksDB, reducing both CPU contention and network traffic. 
GPU acceleration has been successfully applied as well, as shown by gLSM \cite{ToS2024gLSM} that fully exploits GPU parallelism to accelerate compaction.

Apart from single-node acceleration, distributed offloading approaches have emerged as a powerful alternative.
Caas-LSM \cite{SIGMOD2024CaaS-LSM} proposed offloading compaction to remote nodes in distributed LSM-KVS, while FaaS Compaction \cite{CLUSTER2021Supporting} explored using elastic FaaS clusters for compaction tasks. 
Edge computing environments have also benefited from this trend, with Edgepilot \cite{CLOUD2024Coordinating} demonstrating significant write-stall reduction and throughput improvements through compaction offloading in edge federation environments. 
These hardware-offloading approaches are complemented by software solutions like PStore's \cite{TECS2024PStore} asynchronous parallel compaction scheme, which enhances performance by better utilizing available computing resources.

These solutions reflect that LSM-tree compaction bottlenecks require tailored approaches depending on system architecture, workload characteristics, and performance requirements.
While FPGA solutions excel in customizable, high-bandwidth processing, DPUs offer balanced throughput and low latency, and GPUs provide massive parallelism for suitable workloads. 
Similarly, distributed offloading approaches address different scalability challenges in cloud and edge environments. 
The solution continues to develop as researchers seek to optimize the tradeoffs between hardware efficiency, software flexibility, and system complexity.

\textbf{Efficient Use of Hardware.}
Various underlying storage media, including SSD, non-volatile memory (NVM), and persistent memory, are emerging with different capacities and performance levels.
The storage engine of the LSM-KVS can also run on these storage devices, and existing studies used emerging local storage techniques to optimize the design of the LSM-tree.
The first challenge lies in the inefficient combination of LSM-KVS with SSDs due to the inability of current LSM-tree designs to fully exploit the high parallelism capabilities and multi-channel architecture inherent in SSDs, especially when used with NVMe SSDs \cite{VLDB2022treeline,ToS2021Leveraging}.
LOCS \cite{EuroSys2014LOCS} proposed to integrate a customized SSD design with LSM-KVS by exposing internal flash channels to applications, allowing for explicit leverage of the parallel capabilities of SSD.
BlueCache \cite{VLDB2016BlueCache} solved the limitations of DRAM-based KVS (i.e., Memcached and Redis), stores key-value pairs in flash storage, and implements all KVS operations in hardware, including the flash controller.
NoFTL-KV \cite{EDBT2018noftlkv} addressed performance and storage efficiency issues in persistent key-value stores by completely transforming the traditional I/O stack. 
Integrating physical storage management directly into the key-value store architecture, NoFTL-KV eliminates the need for backward compatibility.
Check-In \cite{ISCA2020CheckIn} is an in-storage checkpointing mechanism that collaborates with the storage engine of the host and the flash translation layer of SSDs to efficiently manage data consistency without the drawbacks of conventional mechanisms, such as write amplification and energy consumption, thus addressing the challenges of flash memory management in SSDs.
KVIMR \cite{ATC2021KVIMR} employed a compaction-aware track allocation scheme and a merged RMW approach to enhance performance and achieve higher throughput for deploying LSM-KVS on interlaced magnetic recording-based HDD.
Dotori \cite{VLDB2023Dotori} proposed to enhance functionality and performance by leveraging the unique interface of KVSSDs, with a novel B+-tree design that capitalizes on the characteristics of KVSSDs, thus overcoming the drawback of lower raw device performance than block SSDs.

Several studies combine the write sequential property of LSM-tree and zoned namespace (ZNS) storage \cite{HotStorage2022LLCompaction,HotStorage2022Compaction,TACO2023SplitZNS,ICPADS2023LeanKV,HotStorage2021Iso-KVSSD,TACO2024WA-Zone,TACO2025WA-Zone,ICPP2024Hi-ZNS}.
For example, LSM\_ZGC \cite{HotStorage2020LSM_ZGC} presented a garbage collection scheme for ZNS SSDs that improves upon traditional zone-unit collection by utilizing segment-based garbage collection, group reading, and data segregation into different zones.
Jung et al. \cite{HotStorage2022LLCompaction} solved the fragmentation issue of partially invalidated zones and presented a lifetime-leveling compaction for ZNS SSDs, mitigating space amplification without garbage collection by equalizing the lifetimes of sorted string tables within a zone.
SplitZNS \cite{TACO2023SplitZNS} is an optimization for LSM-tree on ZNS SSDs that improves garbage collection efficiency by introducing small zones.
WALTZ \cite{VLDB2023waltz} is an LSM-KVS designed for ZNS SSDs, leveraging the zone append command for tight tail latency.
WALTZ addresses long tail latency from batch-group writes with ZNS SSD's internal synchronization and introduces WAL zone replacement for fast failover during parallel appends.
It also employs lazy metadata management for lock-free, fast-put queries.

Some prior works proposed to leverage the byte-addressability and persistence characteristics of persistent memory \cite{ATC2022Pacman,VLDB2021revisiting,ICDE2023CacheKV,TOS2023FlatLSM,ICDE2024BushStore}.
For example, NoveLSM \cite{ATC2018NoveLSM} is a persistent LSM-KVS that leverages NVMs for low latency and high throughput. NoveLSM employed a byte-addressable skip list, direct mutability of persistent state, and opportunistic read parallelism to achieve superior performance.
HashKV \cite{ATC2018HashKV} proposed to optimize update performance and reduce write traffic by employing hash-based data grouping for efficient value storage and garbage collection.
NVLSM \cite{ToS2021NVLSM} presented to accumulate data across multiple floors in a logically sorted run, significantly reducing the number of compactions required and the write amplification while minimally impacting read amplification.

In addition, the persistent memory can be used to accelerate critical operations in LSM-KVS.
For example, MatrixKV \cite{ATC2020MatrixKV} proposed to address the write-stall issue caused by the unsorted file in the $L_0$ and the frequent compactions triggered between $L_0$ and $L_1$.
MatrixKV utilized the NVM to perform smaller, cheaper $L_0$-$L_1$ compactions, allowing the compaction process to be completed quickly.
Then, a column compaction method is proposed to realize the fine-grained key ranges and substantially reduce the amount of compaction data.
Pacman \cite{ATC2022Pacman} proposed to offload reference searches, leveraging tagged pointers and DRAM-resident compaction info, redesigning the compaction pipeline, and separating hot and cold objects.
ZigZagDB \cite{ICDE2024ZigZagDB} proposed to integrate DIMM-interface NVM to alleviate write stalls and write amplification.
This hybrid storage approach alternates data access between SSD and NVM in a `ZigZag' pattern, improving write efficiency and space utilization.
Furthermore, the persistent memory is used with block storage devices in tiered configurations to optimize the access of key-value store operations.
For example, Kreon \cite{ToS2021Kreon} is a key-value store designed for flash-based storage servers that addresses CPU overhead and I/O amplification by utilizing partial data reorganization and memory-mapped I/O, thereby reducing the need for compaction and cache usage.
L2SM \cite{ICDE2021L2SM} solved I/O amplification in LSM-KVS and employed a multi-level log structure to segregate and efficiently manage frequently updated key-value items, thereby reducing the disruptive changes to the tree structure and minimizing unnecessary disk I/Os across tree levels.
ChameleonDB \cite{EuroSys2021ChameleonDB} addressed the challenges posed by the hybrid characteristics of Intel's Optane DC persistent memory through combining an LSM-tree structure for efficient, low-write-amplification writes with an in-DRAM hash table for rapid reads.
In addition, Clover \cite{ATC2020Clover} proposed to address the challenge of integrating persistent memory into disaggregated storage systems by proposing a passive disaggregated persistent memory model that enables remote persistent memory management from compute servers.
As non-volatile memory technologies continue to evolve, their tight integration with LSM-trees will remain crucial for building next-generation storage systems that simultaneously achieve low latency, high throughput, and strong persistence guarantees.

\textbf{Heterogeneous Storage.}
With the diversity of storage media and the typically skewed workloads, adopting heterogeneous storage devices to optimize LSM-tree-based systems is a viable and effective solution for LSM-KVS optimizations.
The storage layer can be tailored to better accommodate the access patterns and data distribution of LSM-KVS.
For example, hot data, which is frequently accessed, is placed on faster, higher-performance storage media, while cold data, which is less often needed, is stored on slower, more economical storage media \cite{TC2024SplitDB}.
Heterogeneous storage solutions can help efficiently manage these varying demands, thereby enhancing the overall performance of LSM-tree based applications.
SplitKV \cite{HotStorage2020SplitKV} proposed to enhance compaction operations by tailoring I/O paths for different key-value sizes. Specifically, small key-value items are written directly into persistent memory for subsequent migration to SSDs, while large items are efficiently written to SSDs.
PrismDB \cite{ASPLOS2023PrismDB} and Prism \cite{ASPLOS2023Prism} both leverage multi-tiered NVMe storage architectures, with the former focusing on optimized data migration between 3D XPoint and QLC NAND tiers, and the latter balancing latency and bandwidth across NVMe SSDs and Intel Optane DCPMM while maintaining crash consistency.
Furthermore, some works focused on addressing specific deployment scenarios. 
MirrorKV \cite{SIGMOD2023mirrorkv} solved hybrid cloud storage challenges through a dual-strategy approach: vertical separation of hot/cold data across storage tiers with customized compaction, and horizontal separation using mirrored LSM-trees to optimize caching and locality. 
Chen et al. \cite{ICDE2023Workload} specifically target write amplification reduction in NVM-SSD hybrid systems, employing virtual-split techniques to improve access locality while minimizing additional writes.
These solutions demonstrate how modern LSM-KVS architectures can exploit storage heterogeneity to achieve better performance characteristics.

In conclusion, the application of LSM-trees in large-scale distributed and multi-tenant environments introduces significant challenges in compaction management, resource utilization, and cross-tier optimization.
Distributed key-value store architectures should balance local/remote storage trade-offs, leveraging low-latency local storage for hot key data while offloading colder key data to cost-efficient cloud tiers 
Meanwhile, hardware heterogeneity demands specialized compaction strategies.
ZNS SSDs benefit from zone-aware approaches, while persistent memory enables LSM-KVS designs such as NVM-accelerated compaction and write-amplification reduction.
There are still unresolved challenges, including minimizing cross-node compaction traffic in geo-distributed deployments, achieving tenant-aware resource sharing without isolation overheads, and dynamically adapting data layouts to evolving storage hierarchies.
These challenges are further exacerbated by diverging hardware capabilities across storage tiers and the need to maintain consistent performance under dynamic, skewed workloads.

\subsection{Key-Value Store Operation Optimization}
In this section, we introduce the optimization of key-value store operations, including point lookup, range query, and delete operations.
The existing solutions for key-value store operations are more focused on optimizing the data structure, such as how to achieve fast queries using the Bloom filter.
Also, enhancing the buffer cache hit ratio is a solution to improve the read performance.

\subsubsection{Query Operations}
The optimization of query operations focuses on the order of data between SSTables.
Frequent background compactions can ensure high data orderliness but may have a significant performance impact, leading to write amplification.
The query operation process in LSM-trees involves an initial check of the in-memory data structure, MemTables, to find the latest version.
If the data is not found, the system searches the sorted files and SSTables on disk, starting with the lowest level.
This process will lead to read amplification, that is, reading multiple versions of data from various layers before finding the required key.
The current challenges in query operations lie in optimizing filters, such as point query filters and range query filters, and minimizing read amplification, as well as improving read performance.
Existing work proposed to improve the performance of query operations through the optimization of point lookup and range query, the buffer cache hit ratio, and the key-value separation perspective.

\textbf{Point Lookup and Range Query.}
The LSM-tree often employs point lookup to locate a specific key, while a range query is used to retrieve a set of key-value pairs.
First, to optimize the point lookup, LSM-tree maintains point query filters (i.e., Bloom filters) in memory.
As described in Section \ref{sec:LSM-tree}, each SSTable maintains the data structure of the Bloom filter for quick querying.
However, during point lookup, the required key is not directly hit in the cached Bloom filters and often needs to be checked in the SSTables stored on the disk.
Therefore, the LSM-tree design suffers from read amplification due to multiple Bloom filters check.
Many existing works focused on Bloom filter optimization to improve the query performance.
For example, bLSM \cite{SIGMOD2012bLSM} first proposed using a Bloom filter to optimize read performance.
Monkey \cite{SIGMOD2017Monkey} allocated main memory among Bloom filters to minimize worst-case lookup I/O cost and provide a navigable design space for tuning performance.
TRIAD \cite{ATC2017TRIAD} proposed to optimize the point lookup performance by leveraging Bloom filters to minimize unnecessary I/O operations on popular keys.
ElasticBF \cite{ATC2019ElasticBF} proposed to leverage the data access hotness and proposes a dynamic, fine-grained heterogeneous Bloom filter management scheme to reduce false positives and I/O operations.
Bourbon \cite{OSDI2020from} proposed a machine-learning method to realize fast lookup.
Zhu et al. \cite{DAMON2021Reducing} addressed the CPU costs of Bloom filters in LSM-trees, which are mitigated by the low latency of modern storage devices. They aggressively reuse hash calculations across different Bloom filters and levels, significantly reducing runtime.
Moose \cite{SIGMOD2024Moose} proposed to optimize point lookup, range lookup, and update operations by allowing independent configuration adjustments per level, revealing critical insights for performance optimization.
Furthermore, some works proposed new filter designs, for example, the cuckoo filter that offers lower space overhead than space-optimized Bloom filters \cite{CoNEXT2014Cuckoo,SIGMOD2021Chucky,VLDB2018Morton}.

Second, some scenarios require range queries to find a set of key-value pairs with range query filters.
SuRF \cite{SIGMOD2018SuRF} proposed a data structure that enhances range query filtering in databases by utilizing Fast Succinct Tries, offering improved performance over traditional Bloom filters with tunable false positive rates for both point and range queries.
Rosetta \cite{SIGMOD2020Rosetta} addressed the high false positive rate and performance overhead in range queries of LSM-KVS by employing a hierarchical set of Bloom filters to index all binary prefixes of a key and convert range queries into multiple probes for non-overlapping binary prefixes.
It dynamically adjusts the number of Bloom filters and memory distribution based on workload patterns to optimize the balance between false positive rate and CPU cost.
REMIX \cite{FAST2021REMIX} is a key-value index structure that enhances LSM-tree range queries by providing a globally sorted view across files, enabling fast binary search and retrieval.
Furthermore, Disco \cite{SIGMOD2025Disco} proposed to index all keys across LSM-tree runs to eliminate unnecessary searches and use compact key representations to minimize key comparisons and I/Os, achieving B+-tree-like efficiency for both point and range queries while maintaining low storage overhead.
GRF \cite{SIGMOD2024GRF} addressed CPU bottlenecks from multiple filter probes per query, reduces filter probes to one, and improves performance over current filters by using a novel shape encoding algorithm.
In addition, some work focuses on read performance optimization on secondary index optimization \cite{ATC2023Revisiting,ICDCS2020SineKV,ToS2024Perseid}.

\textbf{Buffer Cache Hit Ratio.}
The LSM-tree buffer, typically implemented as a MemTable and block cache, plays a pivotal role in balancing write throughput, read latency, and compaction efficiency; yet, its management poses unique challenges owing to the log-structured nature of LSM-trees.
First, the MemTable, acting as a write buffer to absorb incoming writes before flushing to disk, faces a trade-off where large MemTables boost write throughput but delay flushes, increasing recovery time and read amplification, which can be optimized through tiered MemTables that separate keys from values to reduce flush overhead.
The block cache, as a read buffer caching frequently accessed SSTable blocks, suffers from compaction-induced invalidation of cached blocks requiring expensive refetches, addressable via compaction-aware caching that predicts post-compaction hot blocks to maintain cache warmth.
A critical insight here is that the MemTable and block cache must be co-optimized, as write buffering decisions directly impact read performance.
Second, flushing triggers latency spikes when a MemTable fills and flushes to $L_0$, temporarily blocking incoming writes, while compaction induces cache pollution by merging SSTables and evicting cached blocks, harming read performance.
Third, adaptive buffer sizing is vital for dynamic workloads, as static sizes lead to either underutilization or contention.
Finally, emerging systems leverage machine learning to predict optimal MemTable flush times, pre-warm caches before compactions complete, and detect workload shifts to adjust caching policies.

To mitigate the performance degradation caused by LSM-tree compaction operations interfering with buffer caching, one possible solution is to increase the buffer cache hit ratio.
By carefully analyzing access patterns and data popularity, the system can prioritize frequently accessed data for caching in the buffer.
In addition, employing a hierarchical multi-level cache architecture, where a smaller and faster cache stores the most critical data and a larger and slower cache stores less frequently accessed data, can further enhance the buffer cache hit ratio.
This tiered cache architecture enables more efficient utilization of available memory resources and minimizes the impact of compaction on read performance.
Several studies \cite{ICDCS2017LSbM-tree} focused on increasing the buffer cache hit ratio by avoiding the buffer resource contention from internal compactions.
LSbM-tree \cite{ICDCS2017LSbM-tree} proposed to add a compaction buffer on disks to minimize cache invalidations caused by compactions.
Some works proposed to increase the cache hit ratio by leveraging the workload characteristics.
AC-Key \cite{ATC2020ACKey} proposed to address read performance degradation in LSM-KVS by introducing an adaptive caching mechanism that dynamically adjusts cache sizes according to workload characteristics.
Some recent works \cite{VLDB2020Leaper, ATC2023calcspar} employed prefetch techniques to improve read performance.
Calcspar \cite{ATC2023calcspar} is a contract-aware LSM-KVS that solves cloud storage's latency variances under different I/O pressures, by regulating I/O request rates and absorbing surplus requests with data cache to reduce tail latency.
Furthermore, leveraging machine-learning models to predict future data access patterns and prefetch relevant data into the buffer can also significantly improve the buffer cache hit ratio, ensuring that the most relevant data is available in the cache when needed.
Leaper \cite{VLDB2020Leaper} proposed to predict and prefetch hot records by machine-learning methods, addressing the cache invalidation issue caused by frequent background operations like compaction and flush.
Leaper achieves higher cache hit rates and reduces access latency, thereby overcoming the limitations of traditional frequency-based cache replacement policies in the presence of mutable record blocks and copy-on-write techniques.

\subsubsection{Delete Operations}
In the design of LSM-trees, the deletion of data does not result in immediate cleanup due to the append-only nature of writes, which logically handle deletions by introducing a $tombstone$ that is only deleted during the compaction process.
Different from regular key-value pairs, the value of the key-value pair in the tombstone is short (commonly one byte).
However, the presence of tombstones also presents challenges.
First, tombstones contribute to storage overhead throughout their lifecycle, persisting in the system until compaction eventually removes them along with their corresponding deleted data entries.
This persistence directly affects storage utilization and operational costs.
Second, and more critically, the physical persistence of deleted data raises privacy concerns, as sensitive information remains recoverable in storage media until compaction processes finally eliminate it. 
The delay between logical deletion and physical removal creates a window of vulnerability where deleted data could potentially be exposed.

These challenges need a trade-off between system performance and privacy guarantee.
On the one hand, frequent compaction ensures that deleted data is removed on time, but it increases write amplification.
On the other hand, infrequent compaction improves write throughput at the expense of extended data durability.
Lethe \cite{SIGMOD2020Lethe} addressed the privacy-related delete issue by reducing write amplification and ensuring timely deletion persistence with a threshold through improved compaction strategies and storage layouts.

\subsection{Discussion}
Although many works contributed to advancements in LSM-tree implementations, several challenges remain that hinder their optimal performance and applicability in emerging architectures and complex application scenarios.
On the one hand, compaction is a potential internal operation that triggers decisions based on various factors, including workload characteristics, underlying LSM-tree designs, hardware features, and the impact of other applications.
As the dataset scales and more storage nodes are required to service key-value store operations, the challenges remain, and the LSM-tree optimization becomes difficult.
This is because (i) the workload characteristics among tens of thousands or even millions of tenants should be identified to provide online and adaptive LSM-tree designs; (ii) the update and iteration of the underlying new hardware provide new solutions for the LSM-tree design; (iii) the distributed architecture involves more impact factors and opportunities.
Therefore, compaction leads to a critical performance impact problem.
On the other hand, key-value store operations often face challenges, including low efficiency in the data structure, such as Bloom filter designs and secondary index structures, a low buffer cache hit ratio due to resource contention from internal operations, and a lack of awareness of user privacy in the delete process.
In conclusion, LSM-tree is everywhere, but there is still room for optimization. 
In the following, we will introduce the optimization design of LSM-KVS from an application perspective in Section \ref{sec:application}, and discuss the advantages and disadvantages of LSM-KVS under different architectures in Section \ref{sec:architecture}.

\section{Application: Real User Scenarios}\label{sec:application}
The challenges of the LSM-tree design also come from the diverse and complex nature of modern applications from different tenants.
For real-time applications, performance is critical, necessitating low latency to accommodate burst writes that occur frequently and unpredictably.
In contrast, high-throughput scenarios such as log archive applications \cite{SIGMOD2021LogStore} propose to leverage numerous batch writes of substantial data sizes, typically executed in offline processes where high throughput is prioritized over low latency.
Considering applications with varying characteristics in modern data centers, a workload-aware LSM-tree design has emerged as a key trend \cite{SIGMOD2023learning}.
This entails dynamically adapting the LSM tree's parameters and behaviors based on the specific workload patterns and requirements, ensuring optimal performance and resource utilization for each distinct application scenario.
This adaptability involves sophisticated workload analysis, real-time monitoring, and intelligent decision-making mechanisms.
Additionally, some applications require a tailored LSM-tree design to maximize performance or meet specific requirements.

\subsection{Workload-aware LSM-tree Designs}
Modern data centers are characterized by dynamic, skewed, and diverse workloads, which necessitate the LSM-tree design to adapt its tuning strategies in real-time to align with the evolving nature of these workloads.
Previously, Cao et al. \cite{FAST2020RocksDBFacebook} proposed to characterize real-world workloads from three Facebook use cases, revealing key insights such as the relationship between key-value sizes and applications, access locality, and diurnal patterns.
They also introduce a key-range-based benchmark to better emulate real-world key-value store workloads, addressing the limitations of existing benchmarks, such as YCSB \cite{YCSB}.
Many existing work studies one important feature of workloads, data locality, to dynamically organize the LSM-tree into different storage layers.
For example, TRIAD \cite{ATC2017TRIAD} focused on improving throughput by leveraging data popularity skew, batching I/O operations, and avoiding duplicate writes through a holistic combination of three techniques.
Mutant \cite{SoCC2018Mutant} is a storage layer for LSM-tree data stores that dynamically organizes SSTables into different storage types based on access frequency, enabling seamless cost-performance trade-offs.
CruiseDB \cite{ICDE2021CruiseDB} proposed to estimate the service capacity of the LSM-tree to adaptively control the number of user requests entering the memory buffer, preventing write stalls and improving SLA (Service Level Agreement) compliance.
In addition, ELECT \cite{FAST2024ELECT} proposed to optimize storage efficiency in skewed workloads through hotness-aware redundancy management. It dynamically converts hot-tier data from replication to erasure coding while optimizing cold-tier offloading, achieving balanced performance across tiers.
RusKey \cite{SIGMOD2023learning} introduced the online orchestration of LSM-tree structures, reinforcement learning-guided transformations, and a new FLSM-tree design for efficient compaction policy transitions, requiring no prior knowledge of the workload.

\subsection{Tailored LSM-tree Designs}
Several works have proposed designing a workload-specific method to optimize the LSM-tree.
Current storage systems optimize for one type of workload at the expense of others due to limitations in data structures.   
For example, VT-tree \cite{FAST2013VTtree} proposed extending LSM-tree to efficiently handle both sequential and file-system workloads, providing a single storage implementation for mixed workloads.
LSM-trie \cite{ATC2015LSM-trie} proposed to reduce metadata and write amplification for managing ultra-large datasets with small data items, achieving high write and read throughput with efficient compaction.
SlimDB \cite{VLDB2017SlimDB} is faster, more memory-efficient for caching metadata indices, and exhibits lower read operation latency compared to existing solutions like LevelDB \cite{LevelDB} and RocksDB \cite{RocksDB}.
In addition, LogStore \cite{SIGMOD2021LogStore} addressed the challenge of managing massive log data in cloud environments with high write throughput and multi-tenancy, proposing a novel cloud-native database that combines shared-nothing and shared-data architectures, utilizes scalable cloud object storage, and employs multi-tenant management and traffic scheduling algorithms to ensure efficient log storage and retrieval.
DiffKV \cite{ATC2021DiffKV} proposed to utilize differentiated management for keys and values, allowing fully-sorted ordering for keys and partially-sorted ordering for values to achieve balanced performance across writes, reads, and scans.
Additionally, machine-learning workloads present challenges for LSM-tree databases due to their requirement for efficient access to large, high-dimensional, and often sparse datasets.
These workloads require rapid data retrieval for both training models and making real-time inferences, which can conflict with LSM-KVS in sequential writes.

\subsection{Multi-tenant Scenarios}

\begin{figure*}[t]
\centering
\begin{minipage}[t]{0.48\textwidth}
    \centering
    \includegraphics[width=1\columnwidth]{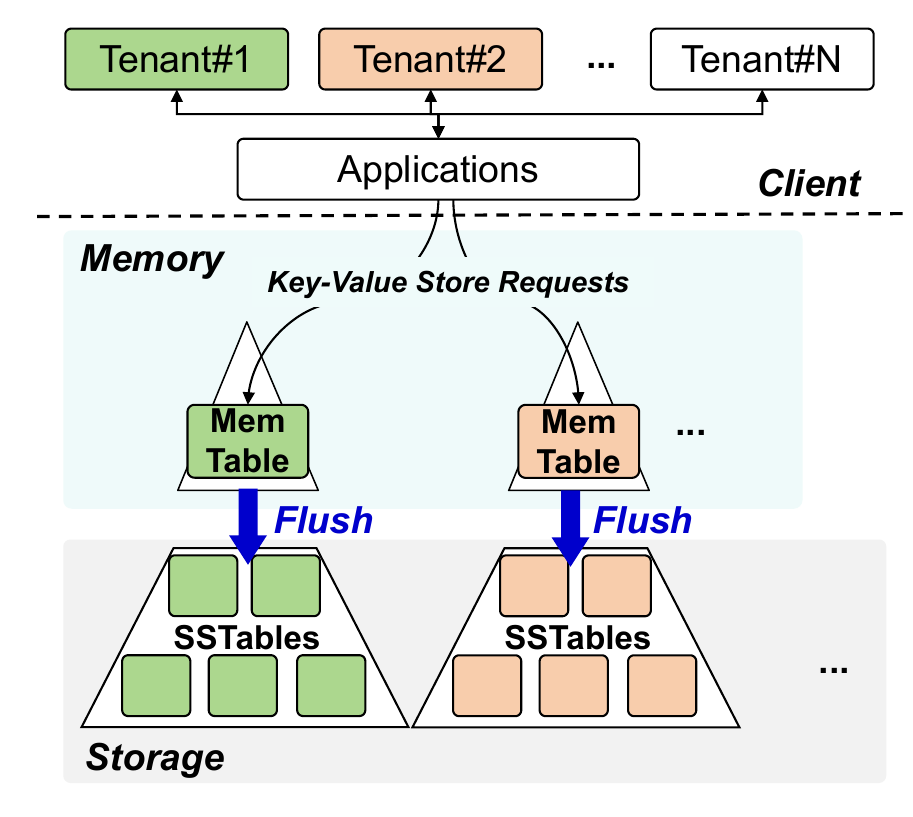}
    \subcaption{Multi-tenant with Separate LSM-tree}
\end{minipage}
\begin{minipage}[t]{0.49\textwidth}
    \centering
    \includegraphics[width=1\columnwidth]{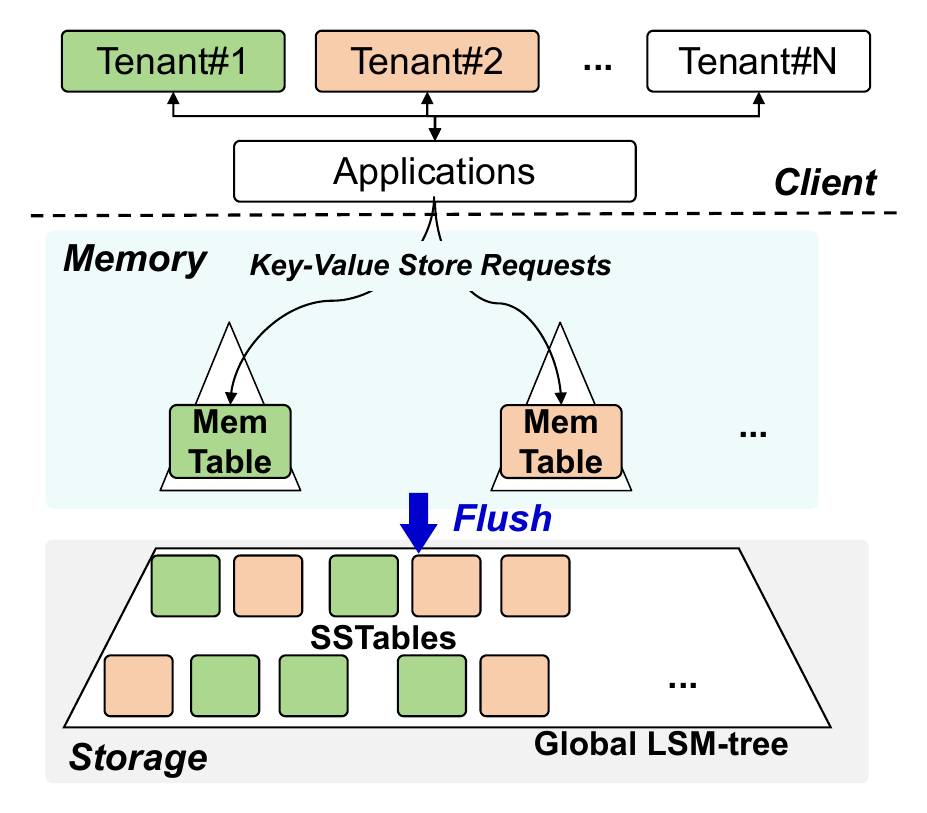}
    \subcaption{Shared LSM-KVS with Logical Separation}
\end{minipage}
\caption{LSM-tree Designs under Multi-tenant Isolation vs. Shared-Storage Scenarios.}
\label{fig:multiTenant}
\end{figure*}

Various applications share disaggregated storage nodes in the cloud environment, necessitating the implementation of multi-tenancy capabilities and ensuring the isolation of distinct tenants.
As the number of tenants increases to tens of thousands or even millions, LSM-KVS must manage the growing amount of data and I/O operations from different tenants.
This leads to increased complexity in compaction processes and performance impacts in different data access patterns.
Figure \ref{fig:multiTenant} illustrates the LSM-tree designs under multi-tenant isolation and multi-tenancy in the shared-storage scenarios.
One solution is resource isolation among tenants for performance consideration \cite{VLDB2022oceanbase,VLDB2023oceanbase,HotStorage2021Iso-KVSSD,TECS2023multitenant}, which ensures that one tenant's performance does not adversely affect others.
This requires a carefully designed LSM-KVS architecture to support multi-tenancy efficiently, including the ability to partition data and resources effectively.
The key challenge of the above solution is that the resources among different tenants are underutilized, and the characteristics of tenants are not considered.
The peak durations of different tenants are usually different, and resources during idle time should be fully utilized while mitigating latency fluctuations during burst workloads.
For resource utilization considerations, an alternative solution is to share storage among multiple tenants.
It allows the system to adapt to the fluctuating resource requirements of various tenants.
However, it is crucial to implement this shared storage solution to preserve I/O isolation and privacy while capitalizing on the advantages of resource pooling.

\subsubsection{Resource Isolation for Performance Consideration}
In the era of cloud computing, LSM-KVS systems often serve data-intensive applications and are often deployed in multi-tenant environments where multiple clients share the same infrastructure.
That is, the multi-tenancy architecture serves multiple LSM-KVS instances of tenants on the same server.
This will lead to one significant challenge: ensuring resource isolation among tenants.
The lack of effective resource isolation can result in performance interference, where the workload of one tenant can negatively impact the performance of others, causing unpredictable latency and throughput. 
Additionally, without proper isolation mechanisms, there is an increased risk of data leakage or unauthorized access across tenant boundaries, which poses serious security concerns.

LSM-tree-based KVSSDs that run the LSM-tree engine in the SSD \cite{ATC2020PinK,MASCOTS2019iLSM-SSD,HotStorage2021Iso-KVSSD}.
Iso-KVSSD \cite{HotStorage2021Iso-KVSSD,TECS2023multitenant} addresses the limitations of namespace and performance isolation in LSM-tree-based KVSSDs for multi-tenant environments by constructing dedicated LSM-trees per namespace and implementing access control based on user namespaces.
Some LSM-KVS systems use a global-shared single LSM-tree to manage the data from multiple tenants \cite{MASCOTS2019iLSM-SSD}, resulting in limited read performance of each tenant.
Some LSM-KVS systems leverage a separate LSM-tree indexing structure for key-value data management \cite{HotStorage2021Iso-KVSSD}.
OceanBase \cite{VLDB2022oceanbase,VLDB2023oceanbase} also allocates separate LSM-tree management for multi-tenants.
The implementation of such resource isolation solutions is crucial for cloud services and key-value stores that serve multiple tenants simultaneously. It enables each tenant to operate within their own dedicated space, free from the interference of others, which is particularly beneficial in scenarios where performance and data security are important.
 
\subsubsection{Shared Storage among Multi-tenants for Resource Utilization Consideration}
In the multi-tenant shared SSD scenario, I/O interference among different tenants is severe, resulting in performance bottlenecks and reduced I/O efficiency. 
This is because multiple tenants are competing for the same physical resources, leading to memory resource constraints, potential SSTable capacity contention, and a decrease in overall performance.
Jaliminche et al. \cite{SoCC2023enabling} addressed the poor performance of workloads when co-located on the same SSD due to I/O Interference, which can lead to violations of Service Level Objectives (SLOs) and low utilization due to excessive overprovisioning.
They involve using machine-learning techniques to predict SSD performance in the presence of interfering tenants, thereby optimizing workload placement.
This overcomes the limitations of previous models by capturing non-uniform workload characteristics and SSD internals, and aggregating features to compute interference among workloads.

\section{LSM-KVS Architecture: From Single-Node to Distributed}\label{sec:architecture}
In this section, we discuss the advantages and potential bottlenecks of LSM-KVS architectures in real-world scenarios.
The typical architecture of key-value stores includes single-node, shared-nothing, and shared-storage.
Table \ref{tab:LSM-KVSarchitectures} illustrates the comparisons of existing key-value store architectures.
In the following, we will introduce these architectures in detail. 

\subsection{Single-Node Architecture}
\begin{figure*}[t]
\centering
\begin{minipage}[t]{0.195\textwidth}
    \centering
    \includegraphics[width=1\columnwidth]{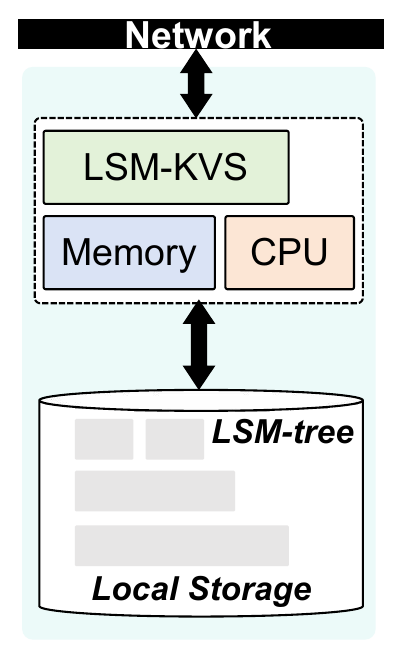}
    \subcaption{Single-Node}
\end{minipage}
\begin{minipage}[t]{0.39\textwidth}
    \centering
    \includegraphics[width=1\columnwidth]{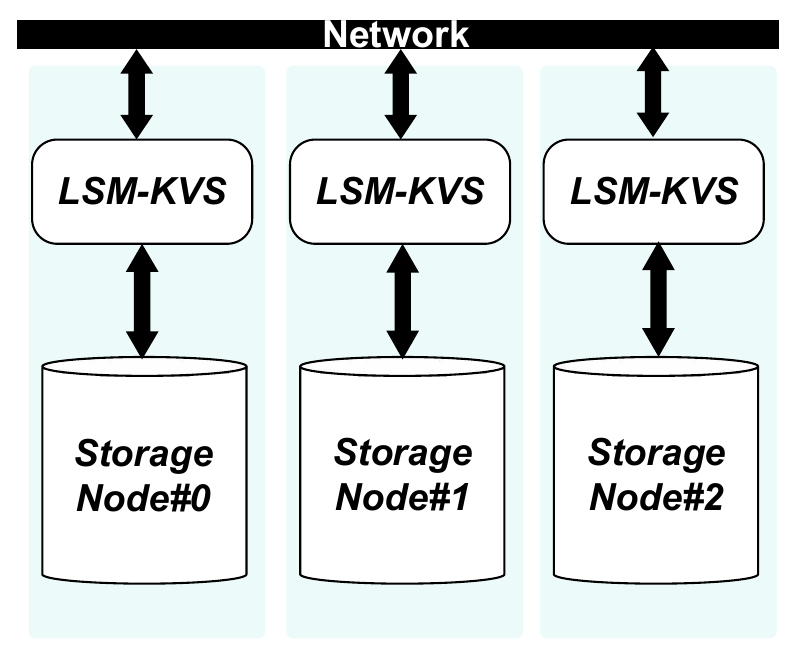}
    \subcaption{Shared-Nothing}
\end{minipage}
\begin{minipage}[t]{0.39\textwidth}
    \centering
    \includegraphics[width=1\columnwidth]{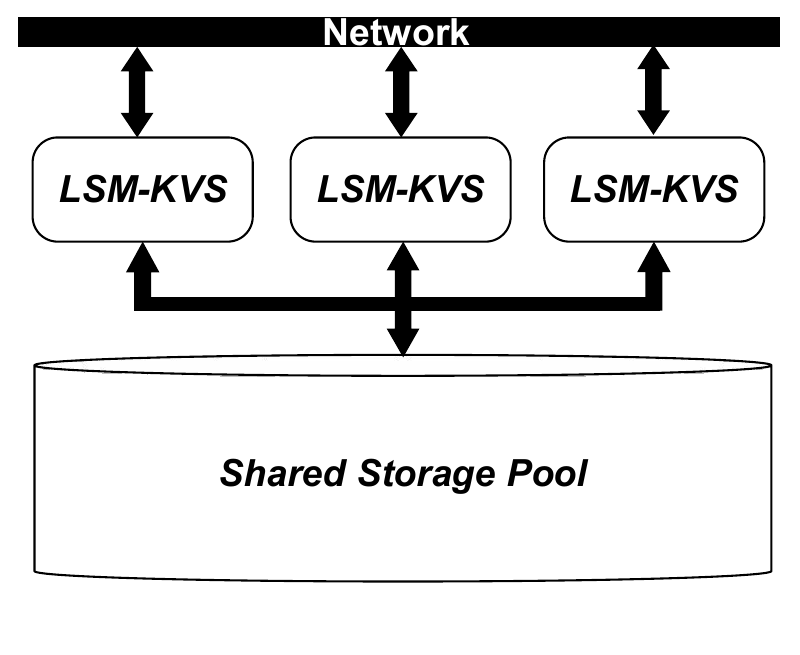}
    \subcaption{Shared-Storage}
\end{minipage}
\vspace{-0.1in}
\caption{Different Key-Value Store Architectures: (a) Single-Node Architecture; (b) Shared-Nothing Architecture; (c) Shared-Storage Architecture.}
\label{fig:architecture}
\vspace{-0.1in}
\end{figure*}

Initially, LSM-KVS is designed for deployment on a single-node server environment, such as LevelDB \cite{LevelDB}, RocksDB \cite{RocksDB}, X-Engine \cite{SIGMOD2019X-Engine}, etc.
As illustrated in Figure \ref{fig:architecture}(a), they are designed to optimize write performance by first writing data to an in-memory structure and then merging it to disk in a sequential manner, which is efficient for write-heavy workloads.
This feature makes them ideal for real-time data generation scenarios.
Additionally, the simplicity and cost-effectiveness of a single-node LSM-KVS allow it to leverage all available resources on a single node without the need for complex distributed infrastructure.
However, this monolithic architecture comes with several limitations.
First, read performance can be compromised, especially for data that has been compacted to disk and has not been accessed recently, as it may require additional I/O operations to merge data from multiple layers. 
The compaction process, which is essential for maintaining performance over time, can be resource-intensive, leading to increased latency and high CPU and I/O resource consumption. This can potentially impact the overall performance of the node during periods of heavy compaction activity.
Second, single-node LSM-KVS has a single point of failure, which means that data redundancy and high availability require additional strategies such as replication or backups. 
Third, this architecture also limits horizontal scalability, making it challenging to expand beyond the capacity of a single node. 
The complexity of recovery after a crash can be a significant issue, as it often involves replaying logs, a time-consuming process.
Lastly, for applications that demand high data locality, single-node LSM-KVS may not be the best fit, as all data is centralized on a single node.

While single-node LSM-KVS offers substantial benefits for certain use cases, including logging, caching, and batch data processing, its suitability must be carefully evaluated against potential drawbacks, particularly in terms of resource contention during compaction, scalability limitations, and single-point-of-failure issues.

\subsection{Shared-Nothing Architecture}
Distributed LSM-KVS designs often employ the shared-nothing architecture, which has evolved from single-node LSM-KVS to the distributed architecture.
As illustrated in Figure \ref{fig:architecture}(b), each node functions independently, with its processing unit, memory, and local storage. This decentralized architecture not only eliminates resource contention but also allows horizontal scaling by integrating additional nodes into the system. 
It excels in scenarios that require high throughput and the ability to process data in parallel.
Examples of shared-nothing architectures include Apache Cassandra and HBase, Google Bigtable and Spanner, Amazon DynamoDB, PingCAP TiKV, Alibaba OceanBase, CockroachDB, Doris, and so on.
Specifically, Cassandra excels in high-availability and scalable environments typical of web applications and IoT. HBase serves big data analytics and real-time queries within the Hadoop ecosystem.
Bigtable supports large-scale analytics and machine-learning workloads, while Spanner is chosen for globally distributed transactions and real-time analytics.
DynamoDB provides fast, managed NoSQL capabilities for mobile, web, and IoT applications.
TiKV combines OLTP and OLAP for financial and real-time analytics.
OceanBase provides high concurrency and availability in financial services.
CockroachDB is a cloud-native SQL database designed for global distribution and e-commerce platforms.
Doris is designed for real-time analytics and data warehousing in advertising and e-commerce.
These databases are selected based on the application's specific needs, including global distribution, consistency, latency, throughput, and data workloads, providing fault tolerance and scalability essential for modern applications.

In the shared-nothing architecture, data is partitioned into many \textit{shards}, with each \textit{shard} being managed by a distinct node.
This segmentation significantly enhances write scalability, as each node operates independently and handles its segment of data, eliminating contention and simplifying scaling.
The failure of one node is isolated and does not impact the system as a whole, ensuring high availability.
However, this architecture closely integrates computing and storage resources, which can lead to significant costs when horizontal scaling and data replication are required during failover recovery.
Also, maintaining data consistency across nodes is a complex task that requires robust mechanisms to ensure data integrity. 
Furthermore, the performance of a shared-nothing system can be sensitive to network latency due to the reliance on inter-node communication for data consistency.
To address these challenges, distributed LSM-KVS solutions often incorporate advanced techniques, such as consistency algorithms, data partitioning strategies for load balancing, and network optimization to minimize latency.
These measures help to maintain the benefits of a shared-nothing architecture while mitigating its potential drawbacks, making it a robust choice for applications that prioritize scalability, performance, and fault tolerance.

\subsection{Shared-Storage Architecture}
Apart from shared-nothing architecture, there is the shared-storage architecture that decouples computing and storage resources.
Based on our research, we have not encountered any LSM-KVS that utilizes a shared-storage architecture.
Here are some possible reasons.
Databases that utilize a shared-storage architecture, such as some relational databases and certain data warehouse systems, do so to simplify data management, enforce strict data consistency, ensure high availability, and optimize for read-intensive workloads. 
These features are particularly beneficial for environments where centralized data control and straightforward recovery processes are essential. 
In contrast, LSM-tree-based databases generally avoid shared-storage architectures because they are designed to optimize for high write throughput, horizontal scalability, and isolated failures.
The write optimization and scalability that LSM-KVS provides can be hindered by the potential contention and latency inherent in shared-storage systems.
In addition, the cost and complexity of maintaining a high-performance shared-storage system that can handle the write loads of an LSM-KVS can be high.
Moreover, ensuring data consistency in LSM-KVS across distributed nodes is inherently complex, and managing this within a shared-storage environment adds further complexity due to the need to coordinate distributed transactions.
As a result, LSM-KVS tends to favor shared-nothing architectures, which allow it to distribute data and processing across multiple nodes without the bottleneck of a central storage system.

In the following, we introduce key-value stores that employ a shared-storage architecture for comparison, although they do not incorporate the LSM-tree structure.
As illustrated in Figure \ref{fig:architecture}(c), multiple compute nodes share access to resources in a storage pool.
Typical examples include Amazon Aurora, Alibaba PolarDB, Microsoft Azure HyperScale, Snowflake, Amazon S3, etc.
Specifically, Aurora is a high-performance relational database service designed for cloud applications that require compatibility with MySQL or PostgreSQL.
PolarDB offers scalability for e-commerce and financial transaction processing.
Hyperscale supports mission-critical SQL databases within the Microsoft Azure cloud.
Snowflake provides a cloud-native platform for large-scale data analytics and sharing.
Amazon S3 serves as a foundational service for backup, archiving, and big data analytics in object storage.
This architecture contrasts with the shared-nothing architecture, where each node has its own local storage.
The storage system is centralized, and databases can be scaled horizontally by adding more compute nodes that share the same storage.
The compute nodes communicate with the storage nodes over a network.
The key benefits of shared-storage architecture include high resource utilization, global data management, etc. 
However, it also introduces challenges related to network overhead and the potential for storage bottlenecks, as all nodes contend for access to the same storage resources.

In conclusion, Table \ref{tab:LSM-KVSarchitectures} lists the comparisons of existing key-value store architectures from single-node, shared-nothing to shared-storage.
Single-node architectures like RocksDB and X-Engine offer simplicity and high write throughput but face limitations in horizontal scalability and fault tolerance.
The shared-nothing systems, exemplified by distributed databases such as Cassandra and TiKV, address these limitations through partitioned processing and storage, achieving superior scalability and availability at the cost of increased complexity in consistency management and resource utilization efficiency.
The shared-storage approach adopted by Aurora and PolarDB represents a middle ground, providing global data management and high resource utilization while introducing new challenges in network overhead and I/O contention.
The architectural choice affects several key aspects: (i) the efficiency of LSM-tree compaction operations, which must adapt to either localized or distributed execution models; (ii) the system's ability to balance workload distribution and resource isolation; and (iii) the complexity of maintaining consistency across different storage layers. 
These trade-offs suggest that future architectures may need to explore hybrid approaches that combine the benefits of these paradigms while mitigating their respective limitations, particularly for large-scale deployments with high performance and availability requirements.

\begin{table*}[t]
\scriptsize
\centering
\caption{Comparisons of Existing Key-Value Store Architectures \tnote{2}. \label{tab:LSM-KVSarchitectures}}
\vspace{-0.1in}
\begin{threeparttable}
\begin{tabularx}{\textwidth}{Xm{3.09cm}m{4.1cm}m{3.8cm}}
\toprule
\textbf{\textcolor{black}{Architectures}} & \textbf{\textcolor{black}{Single-Node Architecture}} & \textbf{\textcolor{black}{Shared-Nothing Architecture}} & \textbf{\textcolor{black}{Shared-Storage Architecture}}\\\midrule
\textbf{Examples} 
&
\begin{itemize}[leftmargin=*]
\item \textbf{Google LevelDB} \cite{LevelDB}
\item \textbf{Facebook RocksDB} \cite{RocksDB}
\item \textbf{Alibaba X-Engine} \cite{SIGMOD2019X-Engine} 
\vspace{-0.1in}
\end{itemize}
&
\begin{itemize}[leftmargin=*]
\item \textbf{Apache Cassandra} \cite{Cassandra}
\item \textbf{Apache HBase} \cite{HBase}
\item \textbf{Google Bigtable} \cite{TOCS2008Bigtable}
\item \textbf{Google Spanner} \cite{OSDI2012Spanner}
\item \textbf{Amazon DynamoDB} \cite{SOSP2007Dynamo}
\item \textbf{PingCAP TiKV} \cite{TiKV}
\item \textbf{Alibaba OceanBase} \cite{VLDB2022oceanbase}
\item \textbf{CockroachDB} \cite{CockroachDB} 
\item \textbf{Doris} \cite{Doris}
\item Huawei GaussDB \cite{GaussDB}
\vspace{-0.1in}
\end{itemize}
&
\begin{itemize}[leftmargin=*]
\item Amazon Aurora \cite{Aurora}
\item Alibaba PolarDB \cite{PolarDB} 
\item Microsoft Azure Hyperscale \cite{HyperScale}
\item Snowflake \cite{Snowflake}
\item Amazon S3 \cite{S3}
\vspace{-0.1in}
\end{itemize}\\\midrule

\textbf{Advantages} 
& 
\begin{itemize}[leftmargin=*]
\item High write throughput
\item Simplicity in deployment
\item High flexibility
\vspace{-0.1in}
\end{itemize}
& 
\begin{itemize}[leftmargin=*]
\item Good write scalability 
\item High availability
\item Isolated failures
\item Simplicity in resource management
\item Parallel processing
\vspace{-0.1in}
\end{itemize}
& 
\begin{itemize}[leftmargin=*]
\item High resource utilization
\item Global data management
\item Simplified data consistency
\item High scalability
\vspace{-0.1in}
\end{itemize} \\\midrule

\textbf{Limitations} 
& 
\begin{itemize}[leftmargin=*]
\item Resource contention during compaction
\item Low parallel proc
\item Low horizontal scalability
\item Single point of failure and complexity in recovery
\vspace{-0.1in}
\end{itemize}
& 
\begin{itemize}[leftmargin=*]
\item Tightly couple computing resources and storage resources
\item Costs of horizontal scaling
\item Data replication overhead
\item Complexity in consistency maintenance
\item Sensitivity to network latency
\item Resource utilization inefficiency
\vspace{-0.1in}
\end{itemize}
& 
\begin{itemize}[leftmargin=*]
\item High network overhead
\item I/O contention issue
\item Resource isolation issue
\vspace{-0.1in}
\end{itemize}
\\\bottomrule
\end{tabularx}
\begin{tablenotes}
\item[2] The LSM-KVS is in bold black.
\end{tablenotes}
\end{threeparttable}
\vspace{-0.15in}
\end{table*}

\section{Research Directions and Future Trends of LSM-KVS}\label{sec:future}
Based on the reviews and discussions of LSM-KVS in Sections \ref{sec:review}-\ref{sec:architecture}, there are still many not-well-solved issues and new challenges for modern LSM-KVS designs, especially in large-scale data environments and dynamic workloads, as well as architecture scalability requirements.
In the following, we list several concerns and provide future trends of LSM-KVS.

\textbf{Trade-off between Query and Write Performance}
Although LSM-KVS is designed for write-heavy applications, the query performance degradation cannot be negligible.
In this case, the trade-off between query and write performance becomes a key research direction.
The key reason for this trade-off lies in the LSM-KVS designs, which prioritize write operations for efficiency.
The compaction process can become a bottleneck, especially if the system has a high write-to-read ratio, as it increases the complexity and time required to serve a query operation.
To mitigate this performance degradation, LSM-KVS implementations often employ compaction strategies.
This not only reclaims storage space but also optimizes the read path by reducing the number of SSTables that need to be scanned during a query. 
However, compaction can be resource-intensive, as it involves significant I/O operations and can temporarily increase latency.
Therefore, effectively managing compactions is key to preserving both fast writes and responsive queries in LSM-KVS systems.
The challenge is to find a suitable balance between write performance and query performance through effective compaction strategies, I/O scheduling, and tuning approaches.

\textbf{Adaptability to Workload Variations.}
In modern data centers and cloud environments, workloads are characterized by their skewness, dynamism, and diversity, which pose significant challenges for traditional LSM-KVS systems. 
The compaction strategies in many LSM-tree implementations are not equipped to handle the rapid and unpredictable shifts in workload demands, often resulting in increased latency and I/O overhead. 
This inflexibility can lead to performance bottlenecks, especially during peak activity periods when the system is under strain.
To tackle this, research is focusing on adaptive compaction strategies that utilize machine learning to predict and adjust to access patterns, as well as on hybrid storage systems that consider data access frequency and store data in different storage mediums.
The goal is to provide LSM-KVS designs that can dynamically optimize compaction and data storage to meet the demands of diverse and evolving workloads without compromising performance.
Another important goal is performance stability, as the workload is dynamically changed and potential compactions.
The accurate predictability of compactions is also necessary.

\textbf{Write and Space Amplification Reduction.}
Write amplification occurs when the same data is rewritten multiple times during the compaction process, which involves merging data from multiple SSTables.
This repeated writing can lead to an increased load on the storage system and reduce the overall lifetime of the storage media, especially in SSDs, where each storage cell has a limited number of program/erase cycles. 
Space amplification, on the other hand, happens when multiple versions of the same data are stored in the LSM-KVS, often as a result of updates or deletions. 
As data volumes increase, driven by AI applications, this can result in significant increases in storage space usage and lead to higher costs and slower query performance.
Reducing write and space amplification in LSM-KVS remains a key research direction.
For example, we can only rewrite necessary data, utilizing data compression to store updates more efficiently, organizing data across storage types to save space, etc.

\textbf{Multi-tenant in LSM-KVS.}
Multi-tenancy architecture in LSM-KVS is crucial for cloud environments, as it enables multiple users to share a single LSM-KVS instance while maintaining data isolation and security. 
This is particularly beneficial for cloud providers and applications with extensive user bases.
However, multi-tenancy presents challenges such as ensuring data security and isolation within shared storage, adapting resource allocation to varying tenant requirements, and maintaining performance while scaling to meet different tenant demands.
To overcome these problems, research directions include creating flexible isolation mechanisms such as virtualization or namespaces, developing adaptive optimizations that adapt to the specific access patterns of each tenant, and studying data partitioning and load balancing to enhance the scalability and resilience of the system in a multi-tenant setting.

\textbf{Scalability in Cloud Environments.}
Scaling LSM-KVS in cloud environments presents unique challenges due to the unpredictable nature of workloads.
Cloud environments are characterized by variability in resource usage, where the demand for computing, storage, and network resources can fluctuate significantly based on user activity and application requirements.
Traditional scaling approaches, such as vertical scaling (upgrading hardware) or horizontal scaling (adding more nodes), can be costly and may not be agile enough to respond to the rapid changes in cloud workloads.
Moreover, these LSM-KVS designs often require manual tuning and configuration to handle varying workloads, which can be complex and time-consuming.
The unpredictability of cloud workloads exacerbates these issues, leading to potential performance bottlenecks if the LSM-KVS cannot scale resources in real-time to match the workload demands.
Additionally, research on more efficient data partitioning and sharding techniques can enhance data distribution and load balancing.

\section{Conclusion}\label{sec:conclusion}
This paper provides a comprehensive discussion of the issues and challenges related with LSM-KVS designs, and then reviews existing solutions, including the design of compaction strategies, optimization of key-value store operations, emerging hardware and architectures, complex applications with dynamic and skewed workloads, and multi-tenant scenarios with diverse requirements.
Moreover, this paper provides research directions and a discussion of future trends for LSM-KVS, including the development of more efficient compaction algorithms, the design of more efficient data structures, the integration of LSM-KVS with cloud-native technologies, and the need for a more robust tuning method to improve performance.
Continued research is imperative to effectively address the challenges of managing LSM-KVS, ensuring the development of adaptive data management solutions that can keep pace with the increasing data volumes, the dynamic evolution of application requirements, and the ongoing advancements in emerging hardware and architecture.

\section{Acknowledgment}
This work was supported by Ant Group through CCF-Ant Research Fund (CCF-AFSG RF20240102).

\bibliographystyle{ACM-Reference-Format}
\bibliography{reference}

% \appendix
\end{document}